\documentclass[10pt,letterpaper]{article}
\usepackage{graphicx}%
\usepackage{dcolumn}
\usepackage{amsmath}
\usepackage{float}
\usepackage{hyphenat}
\begin{document}
\vspace*{0.35in}

%\preprint{qbio/none so far}
\begin{flushleft}
{\Large
\textbf\newline{Fibroblast mediated dynamics in diffusively uncoupled myocytes - a simulation study using 2-cell motifs.}
}
\newline

S. Sridhar \textsuperscript{1,*} and 
Richard.H.Clayton \textsuperscript{1}

\bigskip
\bf{1} Department of Computer Science, University of Sheffield, Sheffield, United Kingdom
\\
\bigskip
* s.seshan@sheffield.ac.uk
\end{flushleft}

\section*{Abstract}
In healthy hearts myocytes are typically coupled to nearest neighbours through gap junctions. Under pathological conditions such as fibrosis, or in scar tissue, or across ablation lines myocytes can uncouple from their neighbours. Electrical conduction may still occur via fibroblasts that not only couple proximal myocytes but can also couple otherwise unconnected regions. We hypothesise that such coupling can alter conduction between myocytes via introduction of delays or by initiation of premature stimuli that can potentially result in reentry or conduction blocks. To test this hypothesis we have developed several $2$-cell motifs and investigated the effect of fibroblast mediated electrical coupling between uncoupled myocytes. We have identified various regimes of myocyte behaviour that depend on the strength of gap-junctional conductance, connection topology, and parameters of the myocyte and fibroblast models. These motifs are useful in developing a mechanistic understanding of long-distance coupling on myocyte dynamics and enable the characterisation of interaction between different features such as myocyte and fibroblast properties, coupling strengths and pacing period. They are computationally inexpensive and allow for incorporation of spatial effects such as conduction velocity. They provide a framework for constructing scar tissue boundaries and enable linking of cellular level interactions with scar induced arrhythmia.

\section*{Introduction}
The heart is an electro\hyp{}mechanical pump, whose coordinated mechanical contraction is enabled by the propagation of synchronized waves of electrical excitation. 
In the mammalian heart, myocytes and fibroblasts constitute two of the most important type of cells, with the larger myocytes being responsible for cardiac electrical activity and the smaller but more numerous fibroblasts maintaining the electro\hyp{}mechanical integrity of the heart~\cite{Camelliti:2005}. Fibroblasts play a crucial role in the repair of heart muscles especially post injury or disease\cite{Camelliti:2005,Kohl:2014}. In aged or diseased hearts the number of fibroblasts may increase substantially (up to $40 \%$\cite{Kawara:2001}) resulting in increased collagen deposition causing fibrosis.

Furthermore fibroblasts themselves differentiate into much larger myofibroblasts in injured and diseased hearts. In this paper we have used the terms fibroblasts and myofibroblasts interchangeably even though we are mostly referring to myofibroblasts. For a long time fibroblasts were not believed to influence the electrical conduction in heart muscles and even today the exact nature of myocyte\hyp{}fibroblast coupling {\it in vivo} is debated~\cite{Kohl:2014,Vasquez:2011}. However several {\it in vitro} experiments have reported the existence of gap\hyp{}junctional coupling between myocytes and fibroblasts under both physiological and pathological conditions~\cite{Gaudesius:2003,Miragoli:2006,Miragoli:2007,Chilton:2007}. Experiments have shown that coupling between myocytes and fibroblasts can significantly alter the conduction properties of the tissue~\cite{Gaudesius:2003,Zlochiver:2008}, excitability of myocytes~\cite{Kizana:2006} and their resting membrane potential~\cite{Miragoli:2006}. At the level of the tissue and organ, fibrosis is known to substantially affect wave propagation in the heart and it is well understood that fibrosis can create a substrate for cardiac arrhythmia~\cite{Tanaka:2007,Nguyen:2012,Morita:2014,Nguyen:2014,Balaban:2018,Campos:2019}.

Several {\it in silico} studies have investigated how fibroblasts alter the electrical activity of individual myocytes and tissue under both normal and pathological conditions~\cite{Jacquemet:2006,Sachse:2008,Jacquemet:2008,Maleckar:2009,Xie:2009,Kazbanov:2016,Sridhar:2017}. These studies have usually simulated fibroblasts as being either attached to a myocyte or inserted in a tissue of coupled myocytes thereby coupling nearest neighbours. However heterocellular cell culture experiments have shown that fibroblasts can enable conduction up to $300\mu m$~\cite{Gaudesius:2003}. {\it In vivo} fibroblasts have been observed to form large sheet-like extensions having additional folds and elongated cytoplasmic processes~\cite{Kohl:2014}. In the sinoatrial node it has been observed that an individual fibroblast could form membrane juxtaposition with a nearby myocytes covering up to $720$ $\mu m^2$~\cite{Kohl:2014,deMaziere:1992}. Fibroblasts that have such long extensions can potentially couple with multiple myocytes that are spatially distant. 

Such long range interactions between distant myocytes mediated via fibroblasts have the potential to modify tissue electrophysiology and dynamics via conduction delays and by exciting resting and partially recovered regions. The possibility of such long range coupling increases in diseased or injured hearts where there are a larger number of myofibroblasts. Ablation lines are another scenario where the fibroblasts that aid the repair of the ablation scars could produce conduction pathways between regions that have been electrically isolated~\cite{Rog:2016}. Another possible scenario occurs when  islands of myocytes are trapped in a scar within a sea of fibroblasts resulting in conduction pathways that connect distant uncoupled regions~\cite{Walker:2007,Kohl:2005}. 

Most computational studies have approached the problem of fibroblast induced dynamics in terms of the effect of the fibroblast distribution, their density~\cite{Kazbanov:2016,Sridhar:2017,tenTuss:2007} and  texture~\cite{Clayton:2018,Jakes:2019}.
On the other hand in this paper we describe a bottom-up approach by considering several simple motifs consisting of $1$ or $2$ fibroblast units coupled to a pair of mutually uncoupled myocytes.The idea of motifs has previously been used successfully to understand networks in complex biological domains including gene transcription, biochemical systems, neuronal networks~\cite{Milo:2002,Song:2005,Alon:2007}. In this paper we have extended the idea of motif to model complex interactions between myocytes and fibroblasts at the scar boundaries.

We hypothesise that fibroblast mediated coupling that connects diffusively uncoupled myocytes can potentially alter conduction between myocytes via the introduction of delays or by the initiation of premature stimuli and can potentially result in reentry or conduction blocks in tissue. We have used the different motifs to investigate the effect of different configurations of fibroblast mediated electrical coupling between mutually uncoupled myocytes. We have also determined the effect of the different motif features such as connection topology, coupling strength, pacing period and cell parameters on the dynamics of the individual myocytes.

The $2$-cell motifs allow for easy comparison of the influence of the intrinsic myocyte dynamics (characterized by its restitution) with the local coupling topology (characterized by the individual motifs) and pacing period. The advantage of using motifs is that they allow for a computationally inexpensive approach to study the effect of individual variation of the many  features of the myocyte-fibroblast coupled system. By incorporating delay in the stimulation times of individual myocytes of the motif, we study the combined effect of conduction velocity in the tissue and the conduction delay arising from the coupling of disconnected myocytes via fibroblasts. Furthermore we have characterized the regimes of myocyte dynamics that can arise from this kind of fibroblast mediated long-range conduction.

\section*{Methods}
The electrical activity of myocytes was described using the $TNNP-TP06$ model of human ventricular cells ~\cite{tenTuss:2004,tenTuss:2006}, while the electrophysiological properties of the fibroblasts were described using the $MacCannell$ ``active" fibroblast model~\cite{MacCannell:2007}. The time variation of the transmembrane voltage $V$ for myocytes coupled to $n$ fibroblasts was given as,
\begin{equation}
     C_m \times \frac{dV}{dt} =  -I_{ion}  + \sum_{k=1}^{n} G_{gap} (V - V_f)
\label{Eq1}
\end{equation}

Here $I_{ion}$ describes the total of all ionic currents:
\begin{equation}
   I_{ion}= I_{Na} + I_{to} + I_{K1} +  I_{Kr}+ I_{Ks} + I_{CaL}+ I_{NaCa} + I_{NaK} + I_{pCa} + I_{pK}
   + I_{bCa} + I_{bNa}
\label{Eq2}
\end{equation}
where $I_{Na}$ is the sodium current, $I_{to}$ is the transient outward current, $I_{K1}$, $I_{Kr}$ and $I_{Ks}$ are the inward rectifier, delayed rectifier
and slow delayed rectifier potassium currents, $I_{CaL}$ is the L-type $Ca^{2+}$ current, $I_{NaK}$ is the $Na^{+}/K^{+}$ pump current, $I_{NaCa}$ is the
$Na^{+}/Ca^{2+}$ exchanger current, $I_{pCa}$ and $I_{pK}$ plateau calcium and potassium currents and $I_{bCa}$ and $I_{bNa}$ are the background $Na^{+}$
and $Ca^{2+}$ currents. $V_f$ is the fibroblast transmembrane potential while $G_{gap}$ is the strength of the gap junctional coupling between myocyte and fibroblast.

The $MacCannell$ fibroblast model equations are used to describe the time evolution of the fibroblast membrane potential $V_f$ (similar to Equation.~\ref{Eq1} and Equation.~\ref{Eq2}) with the ionic currents comprised of inward rectifying potassium current $I_{fK1}$, the time- and voltage -dependent potassium currents $I_{fKv}$,  $I_{fNaK}$ a sodium-potassium pump current and a background sodium current $I_{bNa^+}$.
$C_m$ and $C_f$ are the cell capacitance per unit surface area of myocyte and fibroblast set to $150$pF and $50$pF (corresponding to the larger myofibroblast~\cite{Chilton:2005}) respectively. 

For the myocytes, we used two parameter sets corresponding to {\it Shallow} and {\it Steep} restitution slopes (see Table $2$, slope~$=0.7$ for {\it Shallow} and slope~$=1.8$ for {\it Steep} in ten Tusscher {\it et al}~\cite{tenTuss:2006}).    
The uncoupled fibroblast resting membrane potential $V_{FR}$ were set to either  $-24.5$~mV or $-49.0$~mV~\cite{Nguyen:2012}. Most of the results described in the paper were obtained with $V_{FR}$ set to $-24.5$~mV. A subset of results with $V_{FR} = -49.0$~mV is described in the supplementary material. The different resting membrane potentials were obtained by shifting the gating variable voltage dependence of the time dependent potassium current~\cite{Jacquemet:2008}. 

\begin{figure}[ht]
    \centering
    \includegraphics[width=0.75\textwidth]{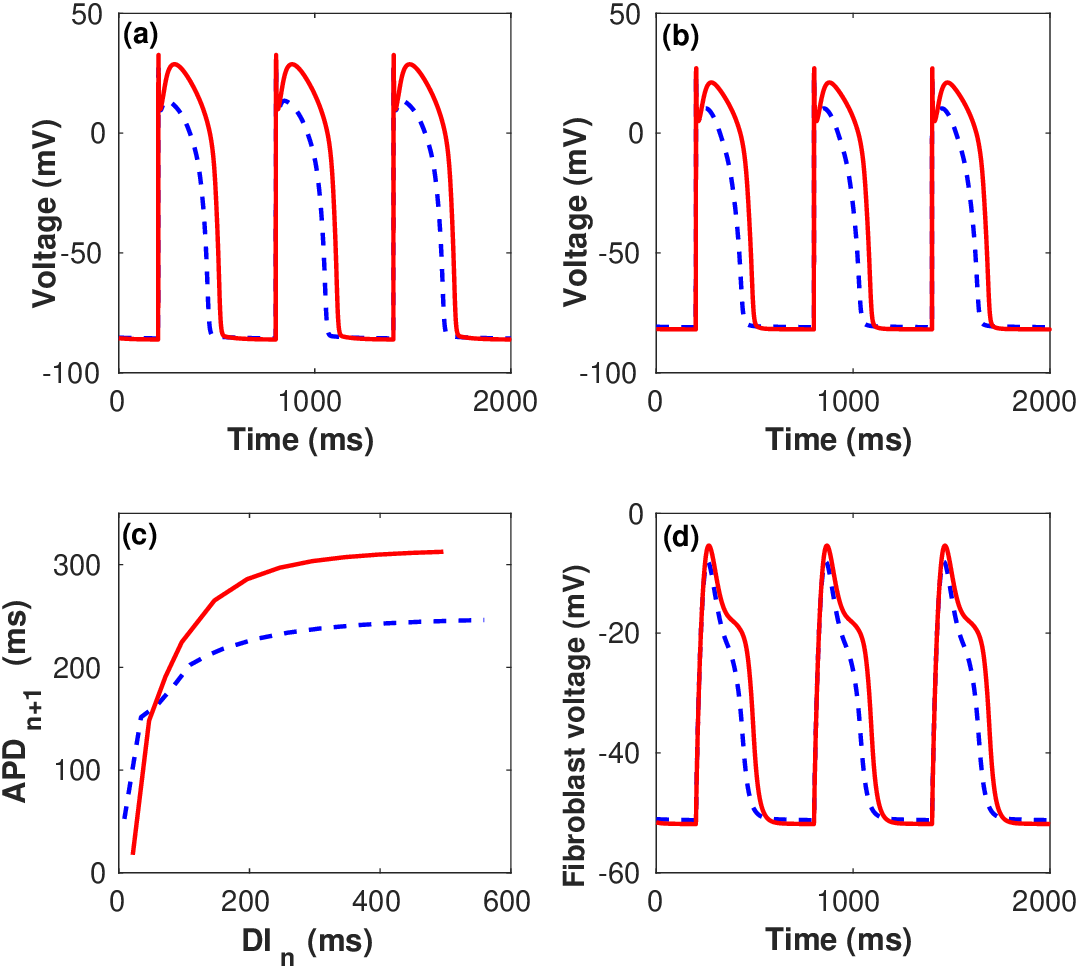}
    \caption{Time series of transmembrane potentials for myocytes and fibroblasts. (a) Transmembrane voltage for uncoupled myocyte (a) and the corresponding S1S2 restitution curves generated at $T = 600$ ms pacing for both {\it Shallow} (broken) and {\it Steep} (solid) parameters (c). The time series of the transmembrane potential of myocyte (b) and fibroblast (resting potential $V_{FR} = -49$ mV) (d) for the case of weak coupling ($G_{gap} =0.5$ nS) for both {\it Shallow} (broken) and {\it Steep} (solid) parameters.}
    \label{fig:MyoAP}
\end{figure}

The action potential of the uncoupled myocyte stimulated at a period of $600$~ms for both {\it Shallow} and {\it Steep} parameters is shown in Fig.~\ref{fig:MyoAP}(a) while Fig.~\ref{fig:MyoAP}(b,d) show the effect of weak fibroblast coupling on the myocyte and fibroblast trans-membrane potentials respectively. In Fig.~\ref{fig:MyoAP}(c) we have plotted the restitution curve for the {\it Shallow} and {\it Steep} uncoupled myocytes.

In our study we considered two identical myocytes coupled either to $1$ or $2$ fibroblast units via different connection motifs (Fig.~\ref{fig:Motif}). Each fibroblast unit represents a fixed number of fibroblasts (see below) connected in parallel and coupled to a neighbouring myocyte (similar to Figure.$1$ in ~\cite{MacCannell:2007} and Figure.$2$ in ~\cite{Kursanov:2023}). 
The different topology considered are (i) $Motif-1$ where one fibroblast unit is coupled to two myocytes, (ii) $Motif-2$ where there is an asymmetry in the number of passive cells connected to each myocyte and (iii) $Motif-3$ where two fibroblast units are coupled in a symmetric fashion with the two myocytes. In the following text we have referred to each fibroblast unit as a single fibroblast ({\it viz.,} Fib $1$ and Fib $2$ in Fig.~\ref{fig:Motif}).
The myocyte-fibroblast interaction is modelled as occurring via a gap-junction. We considered two kinds of gap-junction conductance {\it viz.,} $G_{Loc}$ and $G_{Long}$ corresponding to electrical conduction between fibroblast and the proximal and distal myocytes respectively. Note that for the results reported here the coupling strength of a fibroblast to a proximal myocyte $G_{Loc}$, if a coupling exists between them, is the same for both fibroblast units. Similarly if coupling exists the coupling strength of a fibroblast to a distal myocyte $G_{Long}$ is the same for both myocyte fibroblast units.
For the gap junctional coupling we have chosen values falling in the range ($0-4$ nS) that is considered to be representative of the effect of fibroblasts in cell-cultures~\cite{Jacquemet:2008}.

We captured the effect of conduction velocity in tissue by the introduction of a delay ($\tau_D$) in the stimulation of the distal myocyte ({\it i.e.,} the myocyte coupled to a fibroblast with a coupling strength of $G_{Long}$). While propagation delays of $11 - 68$ ms have been observed in heterocellular culture~\cite{Gaudesius:2003}, in our study we performed simulations with two representative time delays of $\tau_D = 10$ ms and $\tau_D = 25$ ms.

\begin{figure}[ht]
    \includegraphics[width=\textwidth]{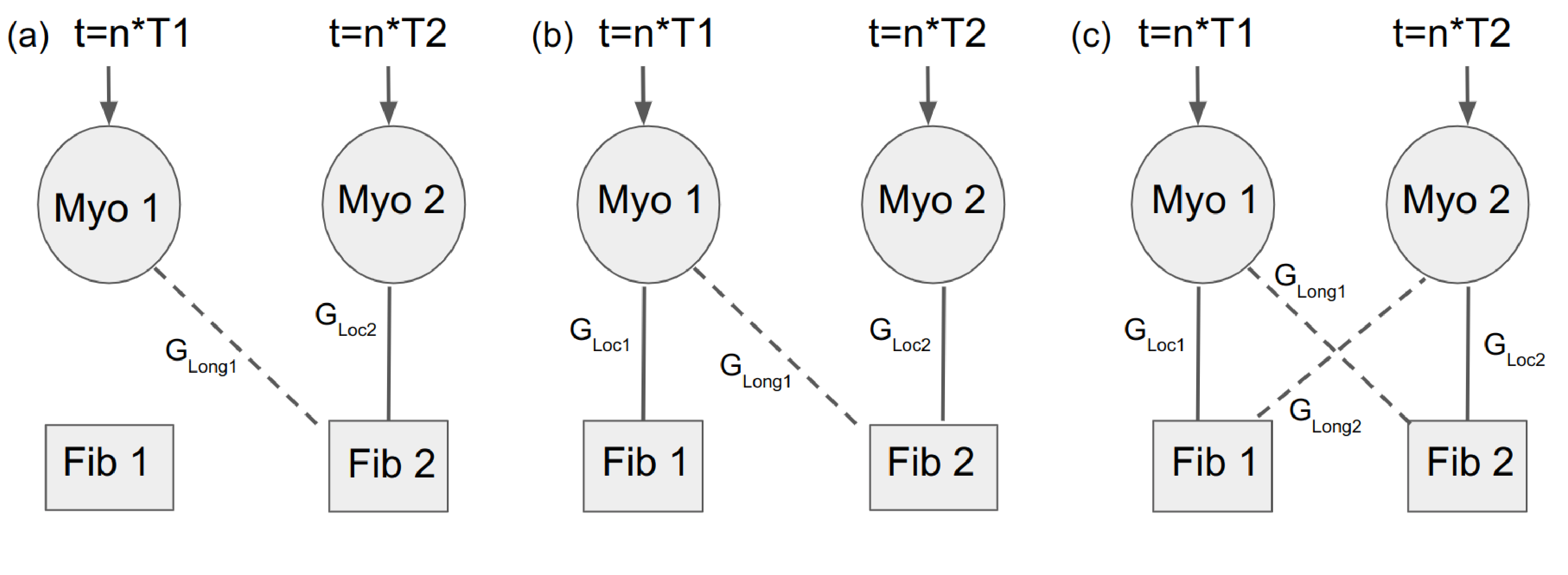}
\caption{Connection topology. The different motifs constructed with two myocytes coupled gap-junctionally to two fibroblast units, {\it viz.,} $Motif-1$ (a), $Motif-2$ (b) and $Motif-3$ (c).}
\label{fig:Motif}
\end{figure}
The resulting equation for the coupled system $Motif-3$ (Fig.~\ref{fig:Motif} (c))
is:
\begin{equation}
     \frac{dV_i}{dt} =  -I_{ion_i} + \frac{N_f}{C_m}\times(G_{Loc_i}\times(V_{f_i} - V_i) + G_{Long_i}\times(V_{f_j} - V_i))
\end{equation}
\begin{equation}
    \frac{dV_{f_i}}{dt} =  -I_{fibion_i} +\frac{1}{C_f}\times(G_{Loc_i}\times(V_i - V_{f_i}) + G_{Long_j}\times(V_j - V_{f_i})) 
\end{equation}
where $i = {1,2}$, $j={1,2}$  and $i \neq j$.
Here $V_i$ and $V_{f_i}$ represent the transmembrane voltage of the $i^{th}$ myocyte and $i^{th}$ fibroblast respectively, while $N_f$ represents the fibroblast density (number of fibroblasts in a fibroblast unit). The equations for $Motif-1$ are obtained by setting $G_{Loc1}$ and $G_{Long2}$ to zero. Similarly $Motif-2$ equations are obtained by setting $G_{Long2} = 0$ nS. 

The coupled ordinary differential equations were solved using an explicit adaptive time step method with a maximum time-step of $0.001$~ms~\cite{Qu1999}. In order to obtain an action potential, either or both the myocytes were individually stimulated every $T$ ms by applying an external stimulus of strength $-52~\mu A/mm^{2}$ for a duration of $1$ ms. While many of the results described in the paper were obtained for $T = 600$~ms, the effect of pacing period was also investigated by setting $T = 500, 400$ and $300$~ms respectively. The fibroblast density $N_f$ was set to $4$. Each simulation was performed by stimulating the myocytes $20$ times, the first $10$ action potentials generated were ignored and the mean action potential duration and fraction of action potentials in the  {\it Pacing} and {\it Response} cells were calculated over the last $10$ action potentials alone. The initial conditions of the variables describing both myocytes and fibroblasts were set to their uncoupled resting membrane values.   
\section*{Results}
\subsection*{Effect of fibroblast coupling on APD}
\begin{figure}[hbt!]
    \centering
    \includegraphics[width=0.75\textwidth]{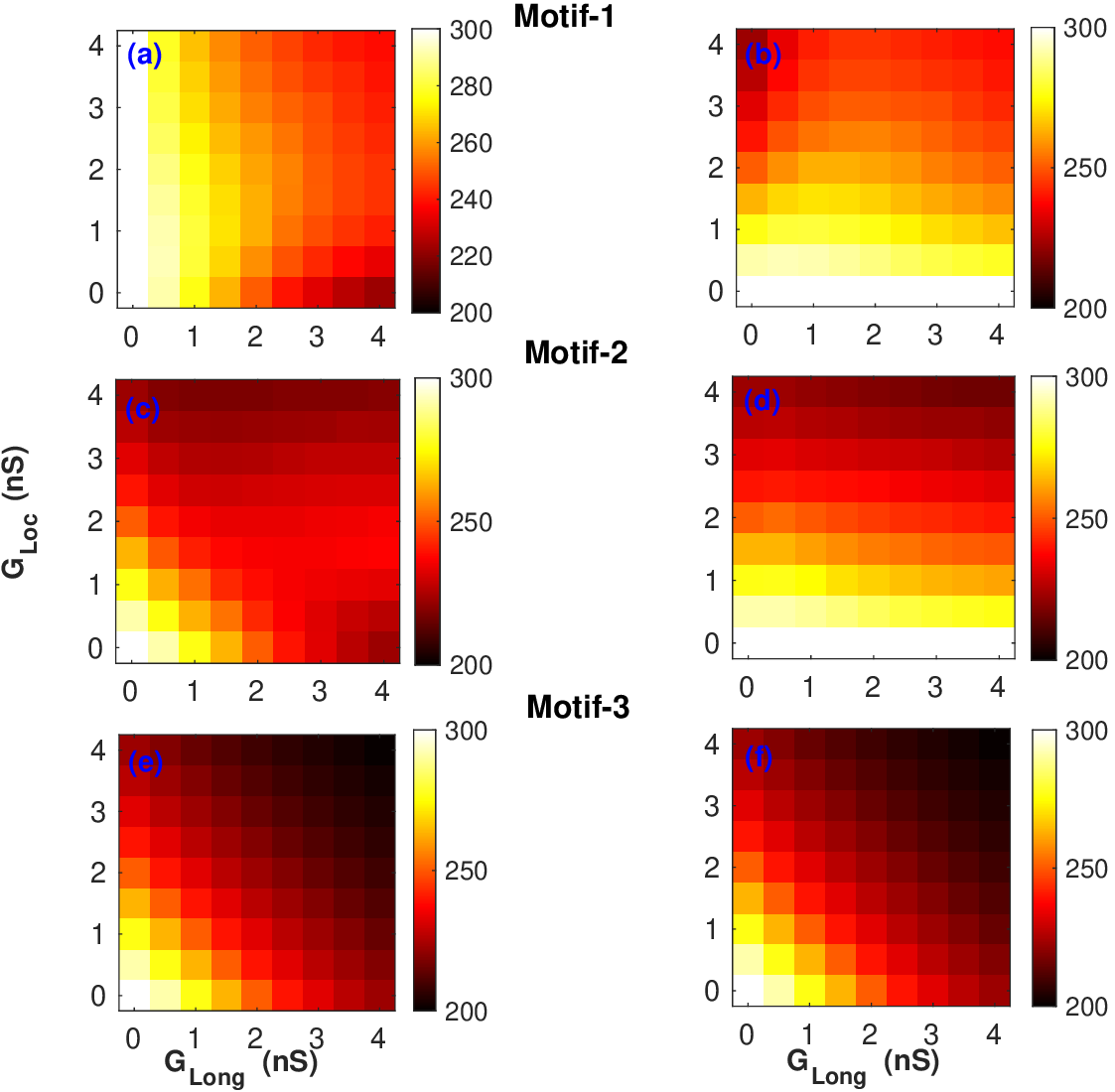}
    \caption{Effect of coupling on APD. Two-parameter plots for all the $3$ motifs describing the effect of the coupling strengths $G_{Loc}, G_{Long}$ on the APD of both $myocyte-1$ (a, c, e) and $myocyte-2$ (b,d,f) with {\it Steep} parameters.}
    \label{fig:APDMotif}
\end{figure}
In order to characterize the effect of connection topology and strength of myocyte\hyp{}fibroblast coupling, we plotted $2$-parameter portraits for the mean APD of the myocytes. The two parameters $G_{Loc}$ and $G_{Long}$ were varied over the range $0-4$ nS in steps of $0.5$ nS.
Figure.~\ref{fig:APDMotif} shows the effect of strength of coupling on mean APD for both $myocyte-1$ (Fig.~\ref{fig:APDMotif} (a,c,e) and $myocyte-2$ (Fig.~\ref{fig:APDMotif} (b,d,f)) for the different motifs with {\it Steep} restitution parameters. (See Supplementary Fig.~\ref{fig:S1} for the equivalent figure for {\it Shallow} parameters). 

While the general effect of fibroblast coupling was to reduce the myocyte $APD$, there was significant variation across motifs. $Motif-1$ was the simplest motif with one fibroblast coupled to two myocytes, and showed a reduction in APD of nearly $70$~ms for the case of the strongest coupling ($G_{Loc} = G_{Long} = 4.0$~nS) as compared to the case of no coupling (Fig.~\ref{fig:APDMotif} (a,b)). However the largest reduction in APD in $myocyte-1$ occurred for the case of ($G_{Loc}, G_{Long}$) $= (0.0,4.0)$~nS (For $myocyte-2$, ($G_{Loc}, G_{Long}$) $= (4.0,0.0)$~nS). 
The reduction in APD for {\it Shallow} parameters was smaller than that of the {\it Steep} parameters, with a maximum reduction of APD compared to that of no coupling being around $44$~ms (see Supplementary Fig.~\ref{fig:S1}). In the case of {\it Shallow} parameters the maximum reduction occurred for the case of ($G_{Loc} = G_{Long} = 4.0$~nS). 

We next considered the effect of APD for $Motif-2$ which has a structural asymmetry with each myocyte exposed to a different number of fibroblasts. Figure.~\ref{fig:APDMotif}(c,d) shows the effect of $Motif-2$ coupling on APD for both $myocyte-1$ and $myocyte-2$ for the {\it Steep} parameters. Due to the asymmetry in their coupling  the magnitude of APD was not the same for both myocytes for a given ($G_{Loc},G_{Long}$) pair. However APD of both myocytes became equal as $G_{Long}$ decreased. The reduction of APD in $myocyte-1$ was greater than in $myocyte-2$. This motif too showed a greater reduction in APD for the {\it Steep} parameters as compared to the {\it Shallow} parameters (Supplementary Fig.~\ref{fig:S1}(c,d)). $Motif-3$ is symmetric in terms of coupling and the reduction in APD for a given coupling strength was the same for both myocytes, i.e. the difference in APD between the myocytes was zero. However the decrease in APD with coupling strength was larger for the {\it Steep} parameter set (Fig.~\ref{fig:APDMotif} (e,f)) compared to the {\it Shallow} parameter case(Supplementary Fig.~\ref{fig:S1} (e,f)). Comparing across the different motifs we find that the reduction in APD with coupling increased with complexity of the topology, with the largest reduction in APD occurring for the case of $Motif-3$.

\begin{figure}[hbt!]
    \centering
    \includegraphics[width=0.75\textwidth]{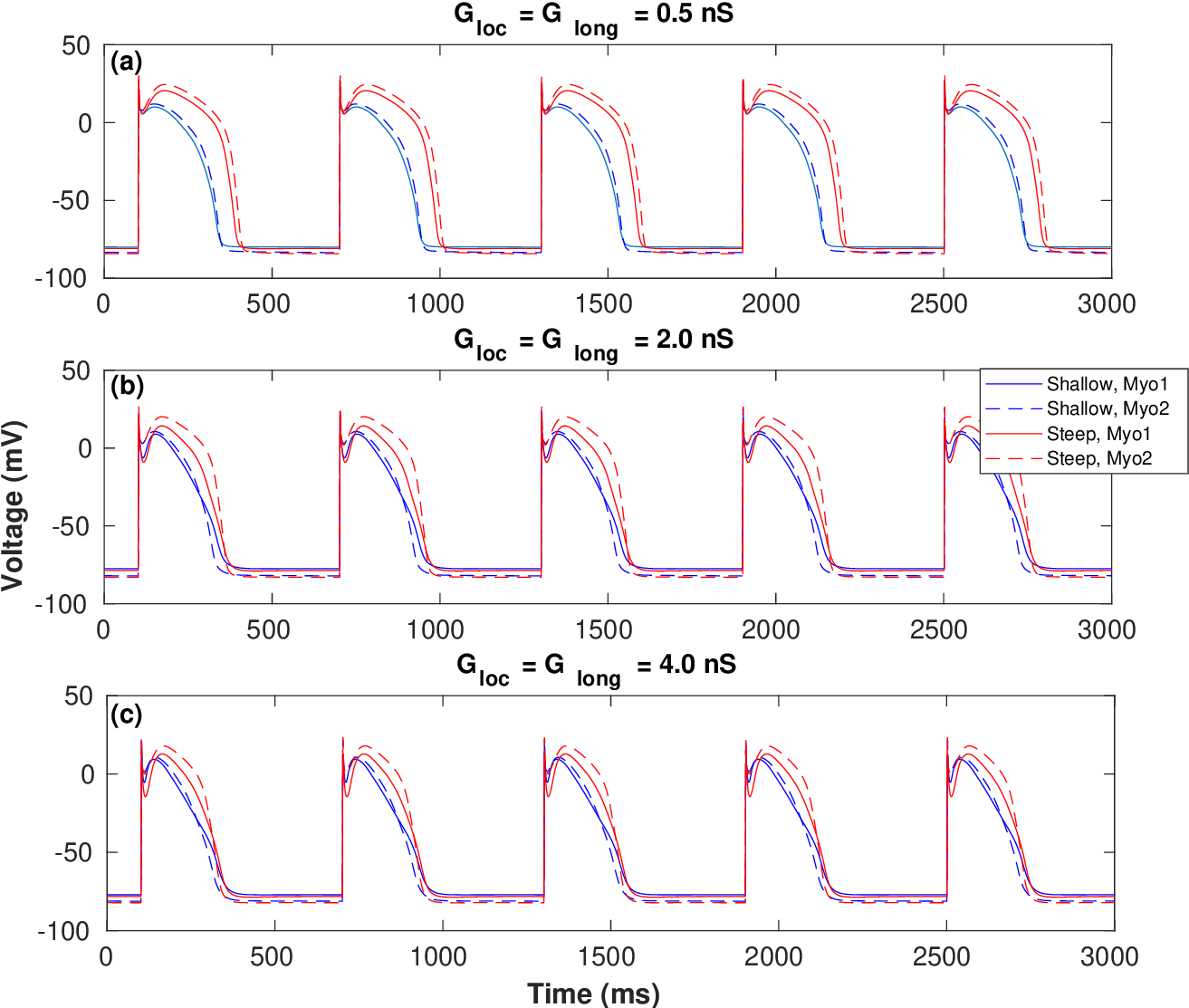}
    \caption{Action potential time series for coupled myocytes. Action potential profiles for both myocytes in $Motif-2$ while pacing the cells at $T=600$ ms at $G_{Loc} = G_{Long} = 0.5$ nS (a), $2.0$ nS (b) and $4.0$ nS (c) respectively. The solid and broken blue lines correspond to $myocyte-1$ and $myocyte-2$ for the {\it Shallow} parameter set while solid and broken red traces correspond to the same for  the {\it Steep} parameter set.}
    \label{fig:APProfile_Motif2}
\end{figure}

For a representative set of coupling strengths we have compared the action potential profiles of both $myocyte-1$ and $myocyte-2$ in $Motif-2$ for both {\it Shallow} and {\it Steep} parameters in Fig.~\ref{fig:APProfile_Motif2}. 
For each of the motifs, there was significant change in features of the myocyte action potential shape including peak, dome and duration of recovery depending on the coupling strengths $(G_{Loc},G_{Long})$.
The corresponding action potential profiles for $Motif-1$ and $Motif-3$ are plotted in the Supplementary section (Fig.~\ref{fig:S2} and Fig.~\ref{fig:S3}).

\subsection*{Effect of delay $\tau_D$ in stimulation}
\begin{figure}[hbt!]
    \centering
    \includegraphics[width=0.75\textwidth]{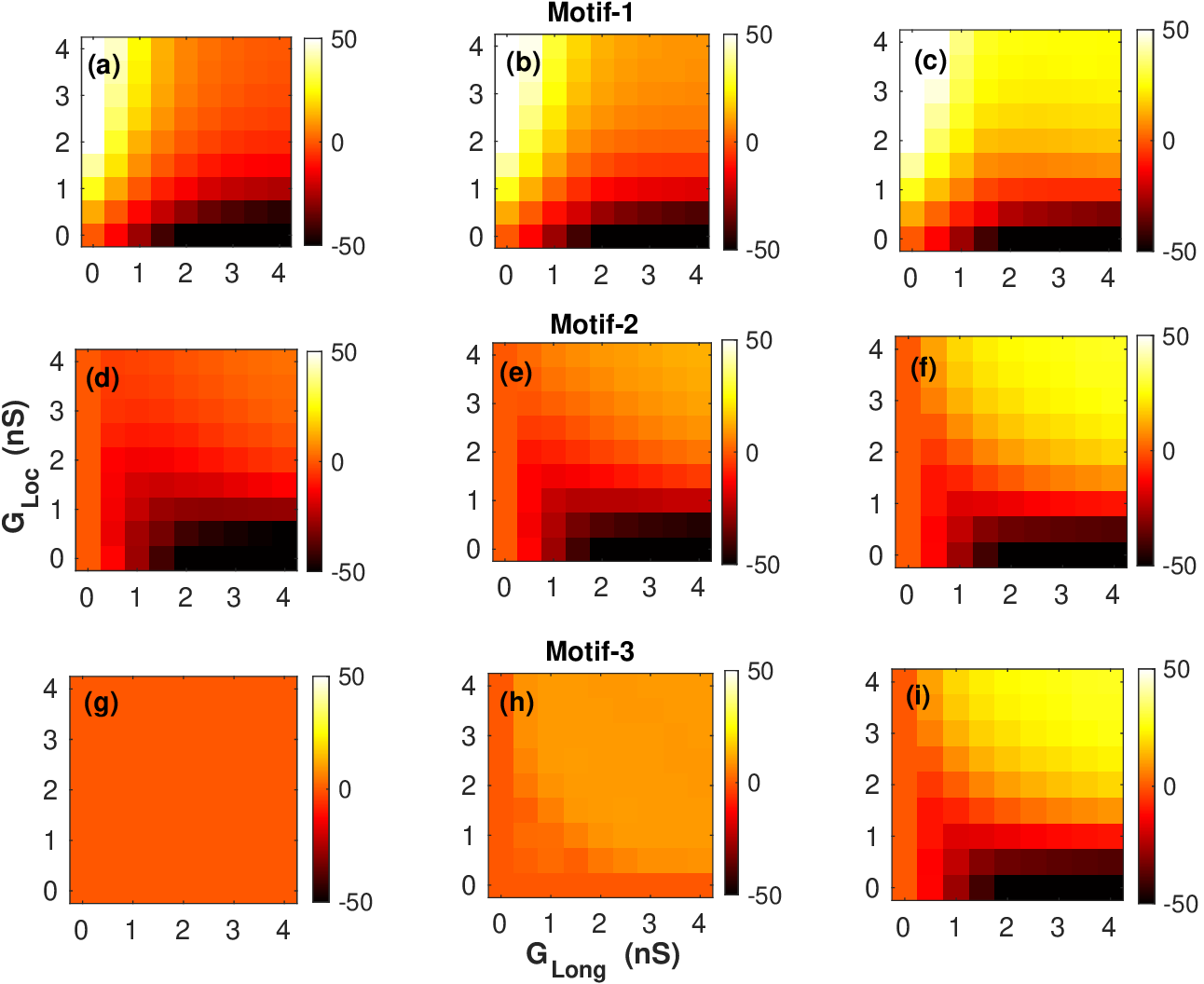}
    \caption{Effect of time delay in the stimulation of $myocyte-2$. The difference in APD ($\Delta APD$) between the two myocytes for $Motif-1$ (a, b, c), $Motif-2$ (d,e,f) and $Motif-3$ (g, h, i) for {\it Steep} parameters. The values of the delay are $\tau_D = 0$~ms (a,d,g), $= 10$~ms (b, e, h) and $= 25$~ms (c, f, i).}
    \label{fig:DelayMotif}
\end{figure}

\begin{figure}[!htb]
    \includegraphics[width=\textwidth]{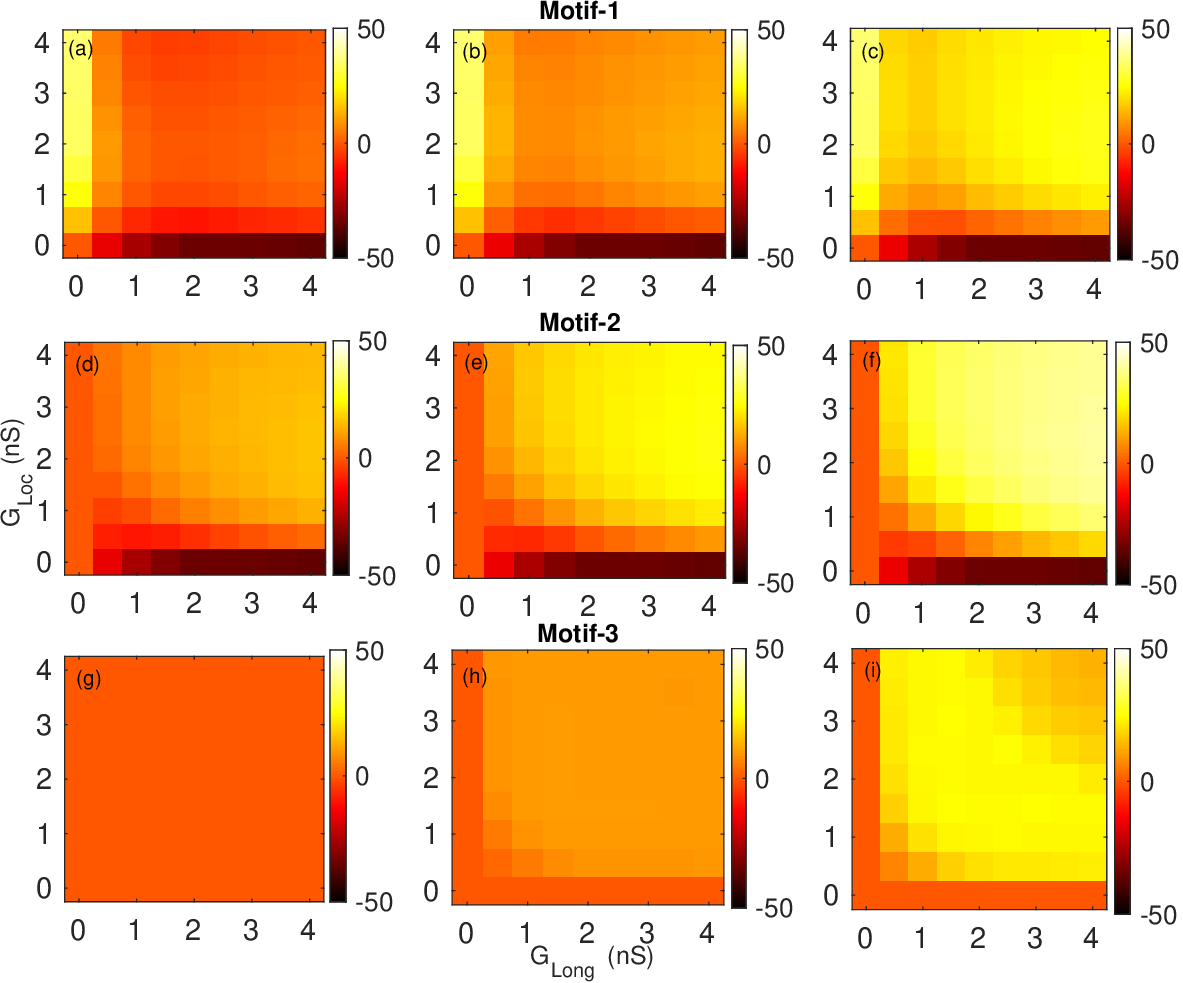}
    \caption{Effect of delay in stimulation of $myocyte-2$ on the difference in the APD of the two myocytes for the different motifs with {\it Shallow} parameters. The values of the delay are $\tau_D = 0$ ms (a,d,g), $= 10$ ms (b, e, h)and $= 25$ ms (c, f, i).}
    \label{fig:S4} 
\end{figure}

In the result described above, both myocytes were stimulated simultaneously. To capture the effect of conduction velocity across tissue we stimulated $myocyte-2$ with a delay after the stimulation of $myocyte-1$. Supplementary Fig.~\ref{fig:S5} describes the time series of myocyte membrane voltage with $myocyte-2$ stimulated with a time delay $\tau_D = 25$~ms. Figure.~\ref{fig:DelayMotif} shows the effect of stimulating $myocyte-2$ with a delay of $0$~ms, $10$~ms and $25$~ms after stimulating $myocyte-1$ on $\Delta APD$, (defined as the difference in APD between $myocyte-1$ and $myocyte-2$). With $Motif-1$, for all values of delay considered here, we observed that for both {\it Steep} 
and {\it Shallow} parameter sets (Supplementary figure Fig.~\ref{fig:S4}(a)) $\Delta APD$ was positive (negative) for $G_{Long} = 0$ nS ($G_{Loc} = 0$ nS). 
However for the {\it Steep} parameters the magnitude of $\Delta APD$ became more negative or less positive (less negative or more positive) with increase of $G_{Long}$ ($G_{Loc}$). On the other hand for the {\it Shallow} parameter set the change in the magnitude of $\Delta APD$ was not monotonic for all values of delay. 

We next considered the effect of delay for the case of $Motif-2$. In the case of zero delay in stimulation for {\it Shallow} parameters $myocyte-2$ showed a greater APD than $myocyte-1$ only for $G_{Loc} \leq 1 nS$. For stronger local coupling ($G_{Loc} > 1$ nS), $\Delta APD$ was positive indicating that the reduction in APD of $myocyte-2$ due to coupling was greater than the reduction in that of $myocyte-1$ (Supplementary figure Fig.~\ref{fig:S4} (d)). However for the case of {\it Steep} parameters with no delay, $\Delta APD \geq 0$ only for $G_{Long} = 0$ or when $G_{Loc}$ was large (Fig.~\ref{fig:DelayMotif}(d)). For small $G_{Loc}$ the reduction in APD was greater for $myocyte-1$. Furthermore the magnitude of $\Delta APD$ was more negative for {\it Steep} compared to the {\it Shallow} parameter set. While for stronger coupling, the magnitude of $\Delta APD$ became more positive for longer delays for both {\it Shallow} and {\it Steep} parameters, with a greater increase for {\it Shallow} compared to {\it Steep} parameters (Supplementary Fig.~\ref{fig:S4} (e,f)). On the other hand for the {\it Steep} case (Fig.~\ref{fig:DelayMotif} (e,f), the reduction in APD of $myocyte-1$ compared to $myocyte-2$ for small $G_{Loc}$ increased with the magnitude of delay. However for all values of delay, $\Delta APD < 0$ ms occurred only when $G_{Loc} \leq 1.5$~nS.
 
In $Motif-3$, for a delay of $0$~ms both the myocytes had the same $APD$ as seen in Fig.~\ref{fig:DelayMotif} (g). For a delay of $10$~ms,
both {\it Steep} (Fig.~\ref{fig:DelayMotif}(h)) and {\it Shallow} parameters (Supplementary figure Fig.~\ref{fig:S4}(h)) show a reduction in APD for $myocyte-2$ (corresponding to a positive value of $\Delta APD$) for all pairs of ($G_{Loc},G_{Long}$).
However for a delay of $25$~ms alone the {\it Shallow} parameter set showed a reduction in $APD$ in $myocyte-2$ for all conductance pairs (Supplementary Fig.~\ref{fig:S4}(i)).
On the other hand for the {\it Steep} parameter set (Fig.~\ref{fig:DelayMotif}(i)) $\Delta APD$ was positive at large ($G_{Loc},G_{Long}$) values, while a large $G_{Long}$ and small $G_{Loc}$ conductance results in a greater reduction of APD for $myocyte-1$ than for $myocyte-2$ (as seen by the negative $\Delta APD$ values). 

\subsection*{Initiating action potentials in a resting cell via fibroblasts}
We next considered the case of infinite delay in stimulating one of the cells. In other words only one myocyte in a motif was stimulated ({\it Pacing Cell}) and the effect of this stimulus on the other myocyte ({\it Response Cell}) via the fibroblast was determined. The goal was to identify regions in the $2$-parameter conductance space that would result in excitation of the quiescent {\it Response Cell}. In this experiment, the {\it Pacing Cell} was stimulated at a fixed period and the effect on the {\it Response Cell} was characterized in terms of the fraction of stimuli in the {\it Pacing Cell} that elicited a response in the {\it Response Cell}. 

\begin{figure}[hbt!]
 \centering
 \includegraphics[width=0.75\textwidth]{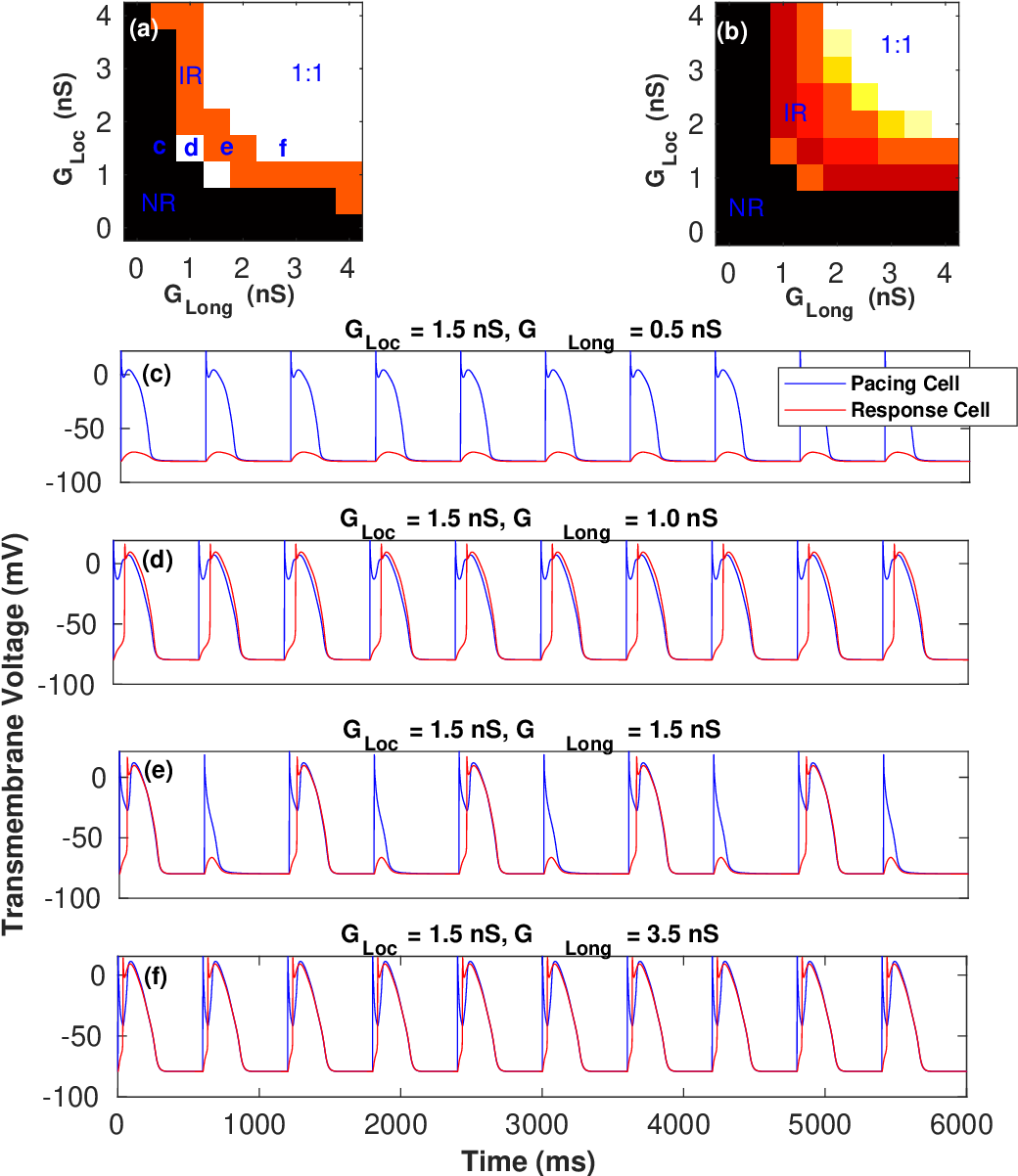}
 \caption{Dynamical regimes described by two-parameter conductance maps. Dynamical regimes characterised as the fraction of stimulus that elicit an action potential in the {\it Response Cell} for both {\it Shallow} (a) and {\it Steep} (b) cell parameters for $Motif-3$. (c-f) The action potential profiles for the parameters marked (c,d,e,f) in the case of {\it Shallow} parameter set (a) comparing the different dynamical behaviour in the Pacing and Response cells.}
\label{fig:APMotif3NoT2}
\end{figure}

In Figure.~\ref{fig:APMotif3NoT2} (a-b), the black region $(NR)$ corresponds to conductance value (small $G_{Loc}$ or $G_{Long}$) that did not produce any excitation in the {\it Response Cell}, even though the {\it Pacing Cell} produced an action potential for every applied stimulus. However for sufficiently strong coupling (larger values of $G_{Loc}$ and $G_{Long}$) every excitation in the {\it Pacing Cell}, was followed by an excitation in the {\it Response Cell}. This parameter region was marked $1:1$. Between $NR$ and $1:1$ regions, there was an intermediate region of parameter values where the {\it Response Cell} did not respond to every excitation of {\it Pacing Cell}. Rather there was an intermediate response $(IR)$, with only some of the action potentials in {\it Pacing Cell} resulting in an action potential in the {\it Response Cell}. 

Figure.~\ref{fig:APMotif3NoT2}(c-f) shows the action potentials for the set of parameter points marked $c,d,e,f$ on Fig.~\ref{fig:APMotif3NoT2}(a). The initiation of excitation at a distal (not directly coupled) myocyte was critically dependent on the ($G_{Loc},G_{Long}$) conductance values in the motif. For $G_{Long} = 0.5$ nS there was no response (Fig.~\ref{fig:APMotif3NoT2})(c). When the value of $G_{Long}$ was further increased to $1.0$, a shorter action potential was elicited in {\it Response Cell} for every stimulation of the {\it Pacing Cell} (Fig.~\ref{fig:APMotif3NoT2})(d). However further increase of $G_{Long}$ to $1.5$ nS gave rise to more complex intermittent dynamics in the {\it Response Cell} (Fig.~\ref{fig:APMotif3NoT2})(e). Finally for $G_{Long} \geq 2.5$ nS, the {\it Pacing Cell} elicited a $1:1$ response in the resting cell (Fig.~\ref{fig:APMotif3NoT2})(f). 

\begin{figure}[!htb]
    \includegraphics[width=\textwidth]{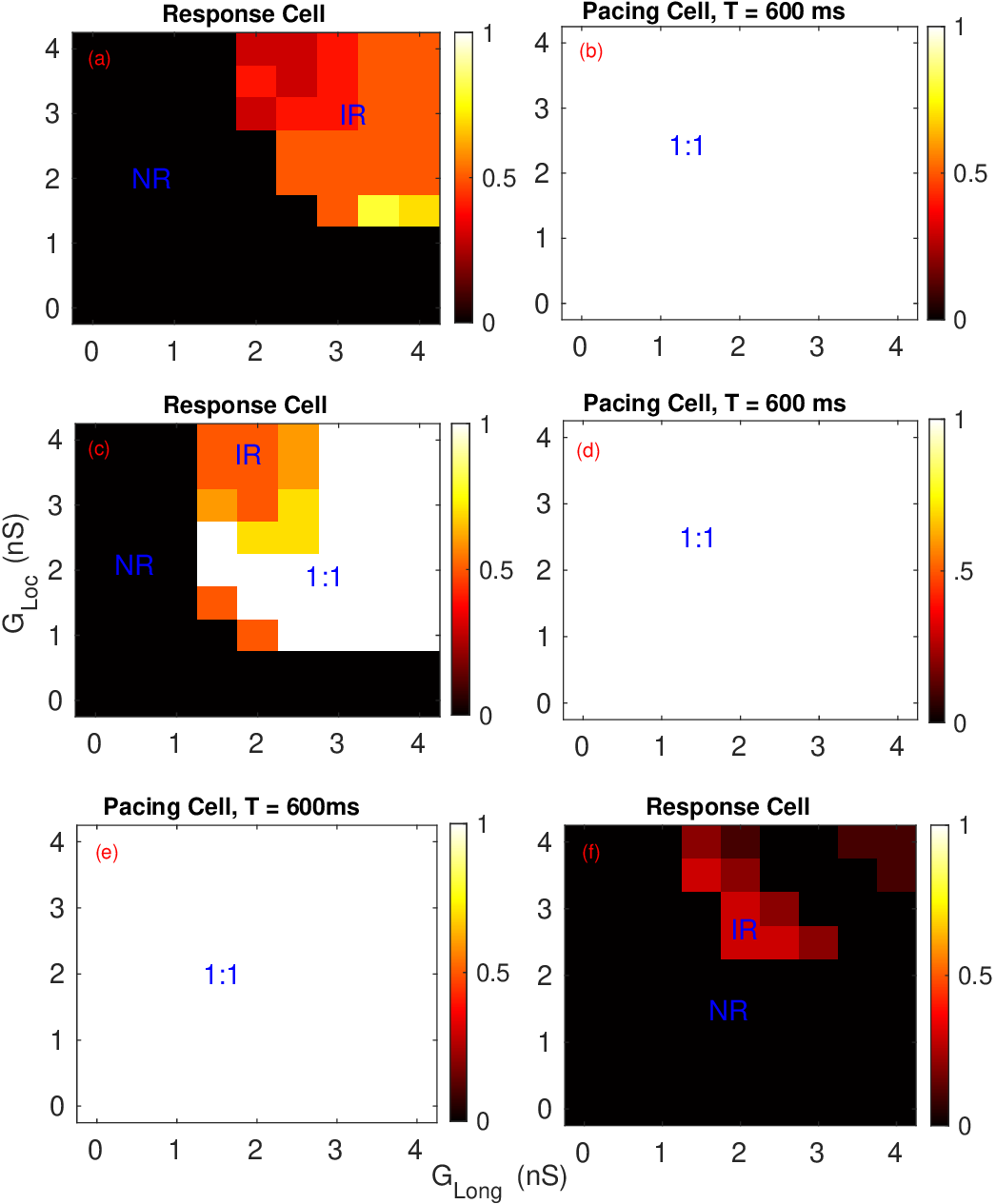}
    \caption{$2$-parameter conductance map for pacing period $ T = 600$ ms describing the different dynamical regimes for the {\it Steep} parameter set, {\it viz.,} $NR$, $IR$ and $1:1$, characterised as the fraction of stimuli that elicit an action potential in the {\it Response Cell} for $Motif-1$ (a,b) and $Motif-2$ (c-f).}
   \label{fig:S6}  
\end{figure}

\begin{figure}[!htb]
    \includegraphics[width=\textwidth]{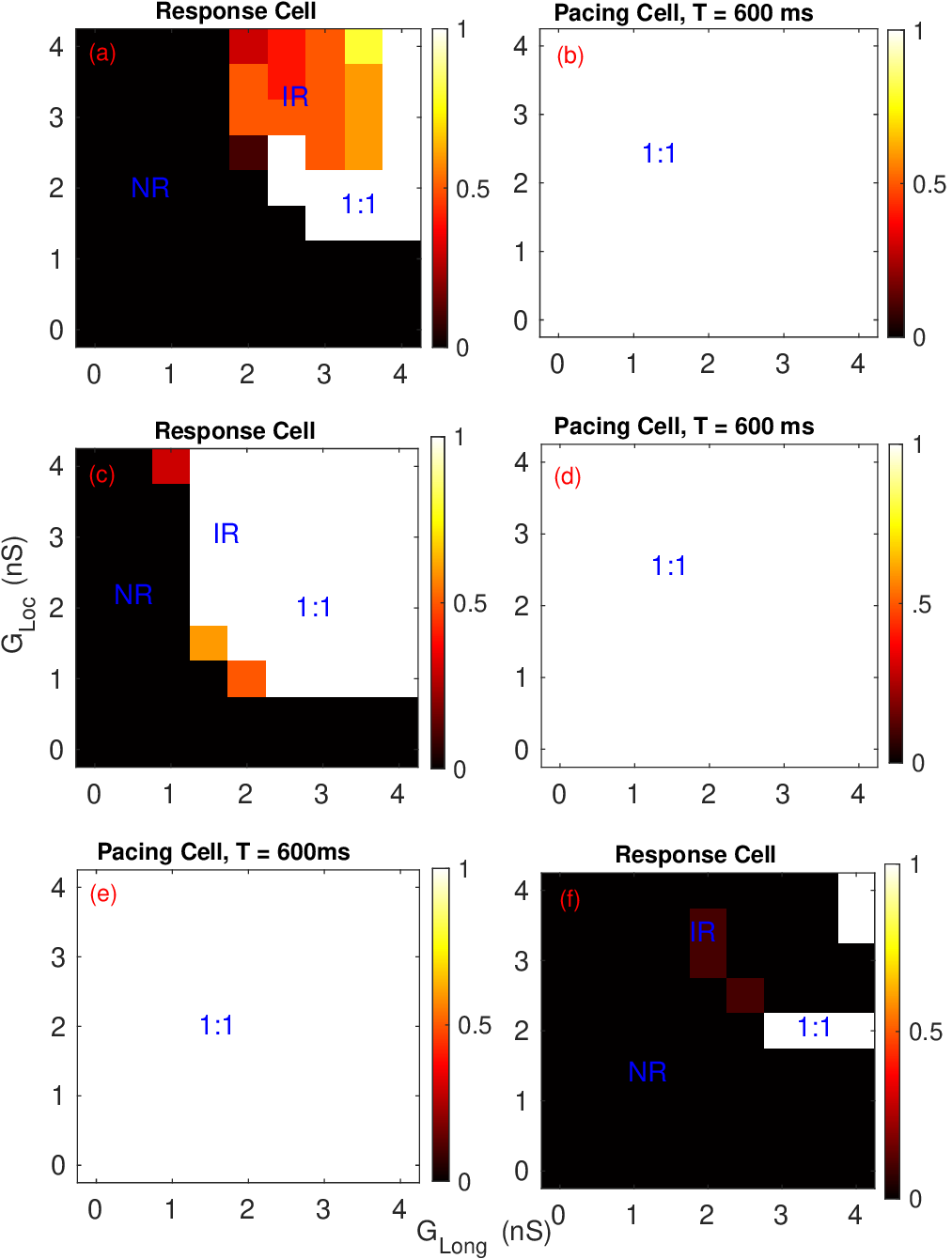}
    \caption{$2$-parameter conductance map for pacing period $ T = 600$ ms describing the different dynamical regimes for the {\it Shallow} parameter set, {\it viz.,} $NR$, $IR$ and $1:1$, characterised as the fraction of stimuli that elicit an action potential in the {\it Response Cell} for $Motif-1$ (a,b) and $Motif-2$ (c-f).}
\label{fig:S7} 
\end{figure}

For the cases where there was a depolarization in the {\it Response Cell}, there was always a time delay with respect to the depolarization of {\it Pacing Cell}. However this delay was a function of the conductance strength (as seen in (Fig.~\ref{fig:APMotif3NoT2})(d,f). 
The action potential profiles in (Fig.~\ref{fig:APMotif3NoT2})(d,f) suggest a reduction in APD of the {\it Response Cell} compared to the {\it Pacing Cell}. Furthermore for large conductance there was an overlap in the repolarization time-series of both myocytes suggesting a synchronization of the recovery process in both the myocytes. Similar $2$-parameter plots for the {\it Pacing} and {\it Response} cells are plotted for both $Motif-1$ and $Motif-2$ (Supplementary figures Fig.~\ref{fig:S6} and Fig.~\ref{fig:S7}). 

Since $Motif-2$ had an asymmetry, we considered two scenarios, {\it viz.,} stimulating either $myocyte-2$ (Supplementary figures Fig.~\ref{fig:S6}(c,d) and Fig.~\ref{fig:S7} (c,d)) or $myocyte-1$ (Supplementary figures Fig.~\ref{fig:S6}(e-f) and Fig.~\ref{fig:S7}(e-f)). Stimulating only $myocyte-2$ could initiate action potentials in $myocyte-1$ for a large range of conductance for both {\it Steep} and {\it Shallow} parameter sets. While for both {\it Steep} and {\it Shallow} parameters, small $G_{Loc}$ or $G_{Long}$ values did not produce a response in $myocyte-1$, at larger conductance $1:1$ response was obtained for both the parameter sets. The region of intermediate response $(IR)$ was larger for {\it Steep} compared to {\it Shallow}. On the other hand a very small number of conductance pairs could initiate action potential in $myocyte-2$ when $myocyte-1$ alone was stimulated for the {\it Shallow} parameter set. For the {\it Steep} parameter set, $1:1$ response was not observed for any of the conductance values. 
  
\begin{figure}[!htb]
    \includegraphics[width=\textwidth]{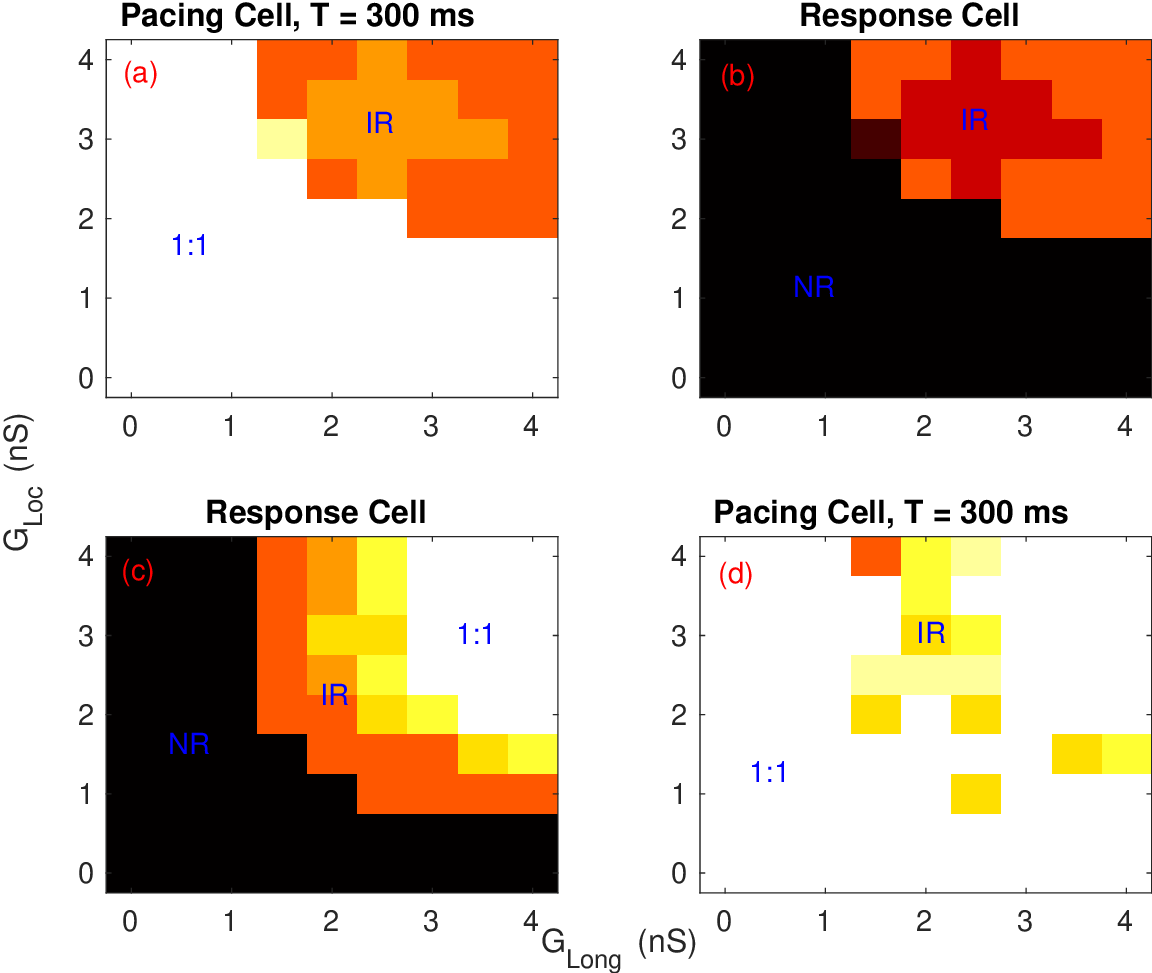}
    \caption{$2$-parameter conductance map describing the different dynamical regimes for the {\it Shallow} parameter, {\it viz.,} $NR$, $IR$ and $1:1$, characterised as the fraction of stimuli that elicit an action potential in the {\it Response Cell} for $Motif-2$ at pacing period $T = 300$ ms.}
 \label{fig:S8} 
\end{figure}

\begin{figure}[!htb]
    \includegraphics[width=\textwidth]{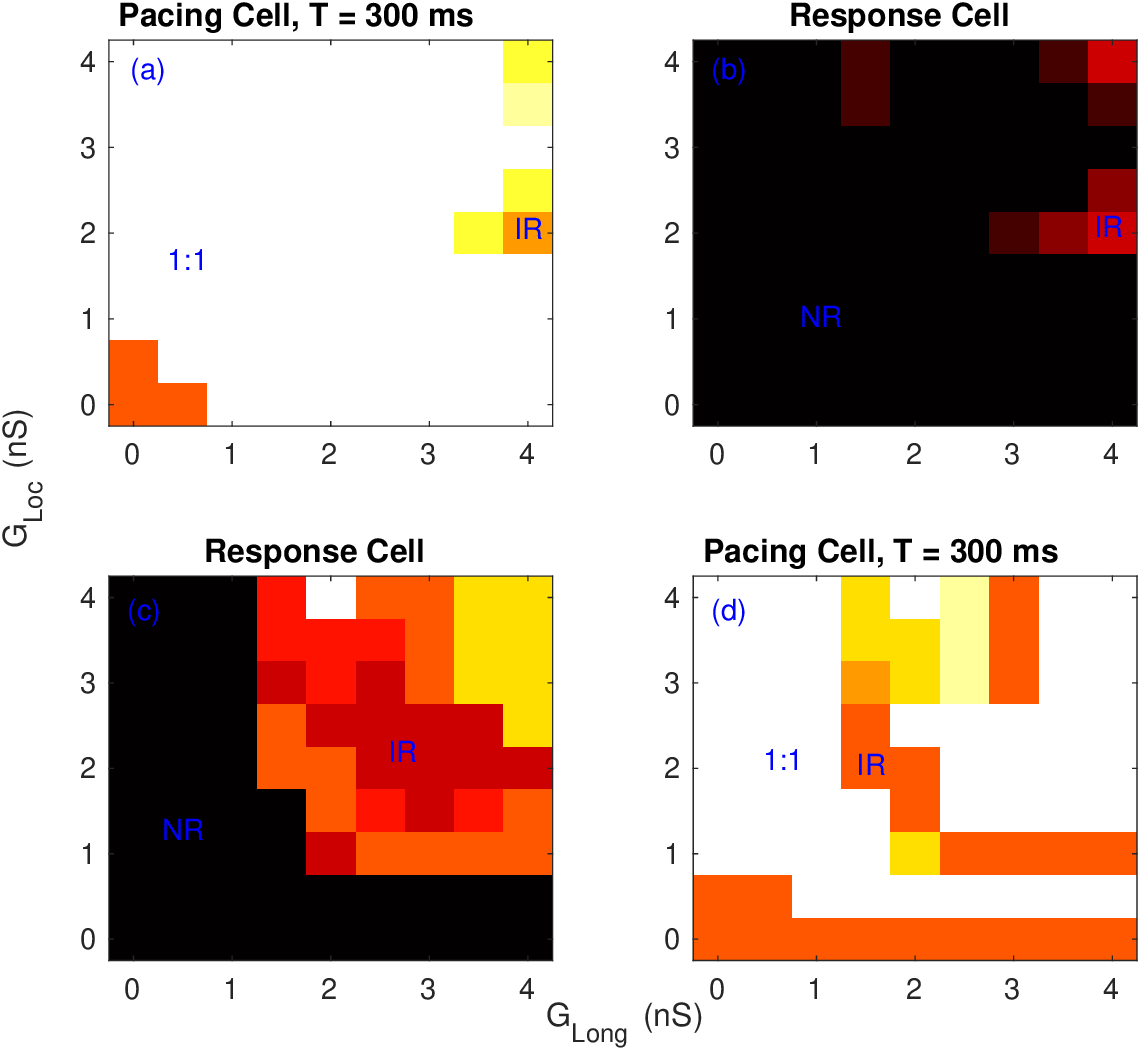}
    \caption{$2$-parameter conductance map describing the different dynamical regimes for the {\it Steep} parameter set, {\it viz.,} $NR$, $IR$ and $1:1$, characterised as the fraction of stimuli that elicit an action potential in the {\it Response Cell} for $Motif-2$ at pacing period $T = 300$ ms.}
\label{fig:S9} 
\end{figure}

\begin{figure}[!htb]
    \includegraphics[width=\textwidth]{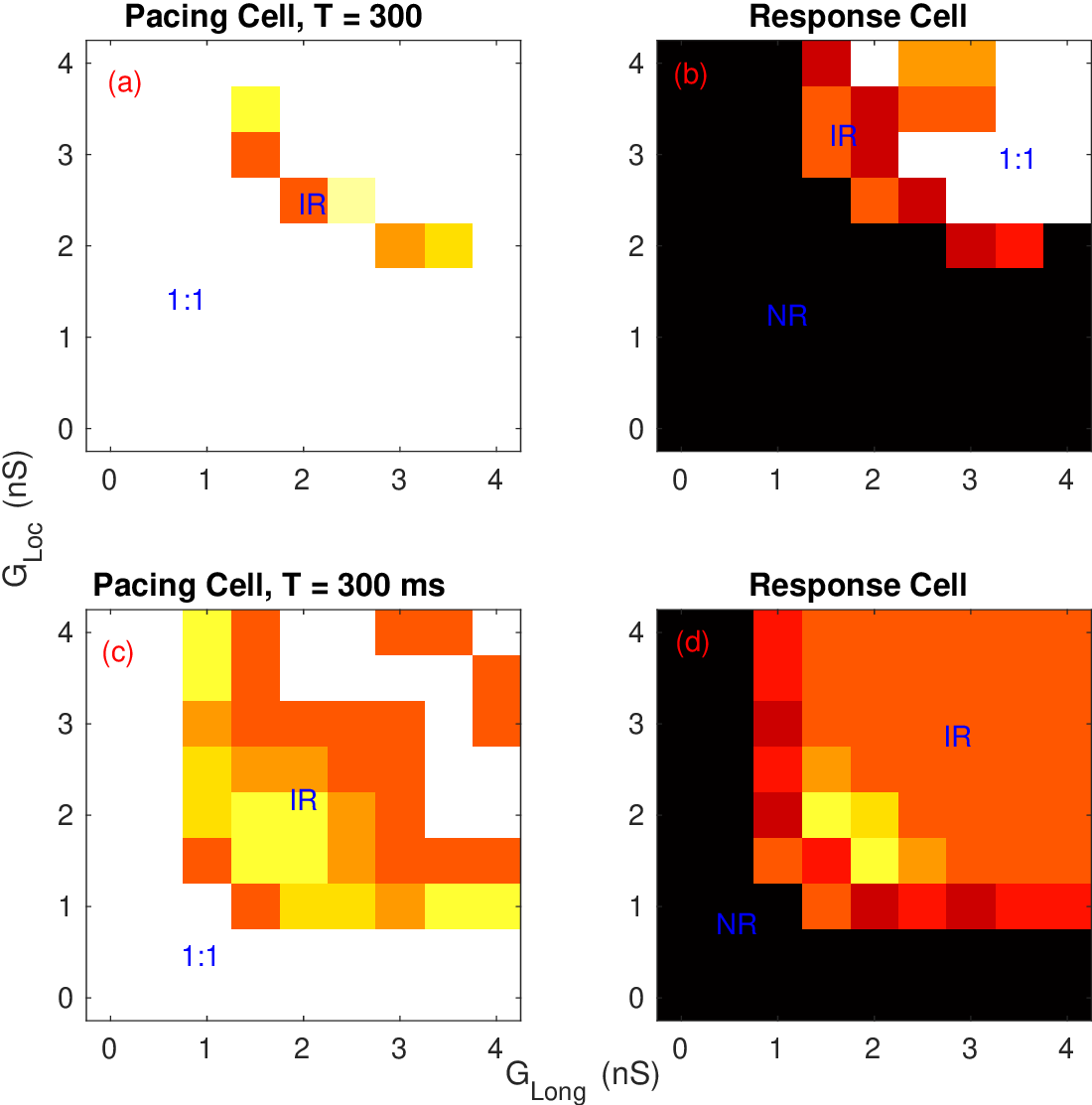}
    \caption{$2$-parameter conductance map describing the different dynamical regimes for the {\it Shallow} parameter, {\it viz.,} $NR$, $IR$ and $1:1$, characterised as the fraction of stimuli that elicit an action potential in the {\it Response Cell} for $Motif-1$ (a,b) and $Motif-3$ (c,d) at pacing period $T = 300$ ms.}
\label{fig:S10} 
\end{figure}

\begin{figure}[!htb]
    \includegraphics[width=\textwidth]{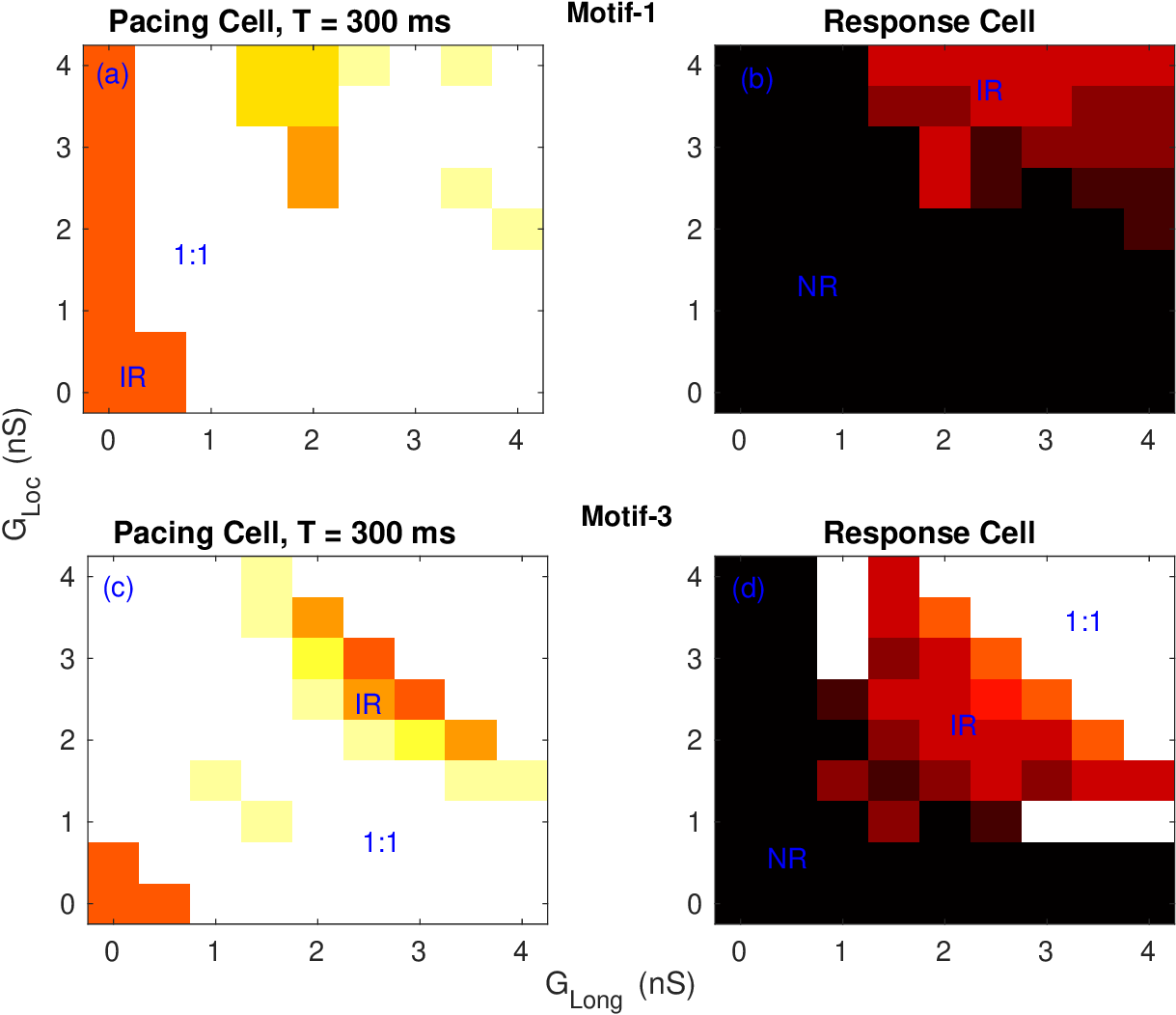}
    \caption{$2$-parameter conductance map describing the different dynamical regimes for the {\it Steep} parameter set, {\it viz.,} $NR$, $IR$ and $1:1$, characterised as the fraction of stimuli that elicit an action potential in the {\it Response Cell} for $Motif-1$ (a,b) and $Motif-3$ (c,d) at pacing period $T = 300$ ms.}
\label{fig:S11} 
\end{figure}

In order to identify the effect of the pacing period on the initiation of action potentials via fibroblasts in a resting cell, we performed the above simulations for pacing periods $T = 500$~ms,
$T= 400$~ms and $T = 300$~ms. The parameter space diagrams for $T = 500$~ms and $T = 400$~ms (not shown) were found to be qualitatively similar to those obtained for $T = 600$~ms, with only the boundaries between the different regimes and size of regimes varying marginally depending on the pacing period. On the other hand pacing at $T = 300$~ms produced regimes that were spatially more patchy suggesting sensitive dependence to coupling strength at very rapid pacing. Furthermore at $T = 300$~ms pacing $IR$ regimes were also observed in the {\it Pacing Cell} implying that at very rapid pacing $1:1$ response was not always guaranteed especially when coupled to other cells that can act as a current sink. For $T = 300$, supplementary figures Fig.~\ref{fig:S8} and Fig.~\ref{fig:S9} describe the different dynamical regimes obtained by stimulating either $myocyte-1$ or $myocyte-2$ for the case of $Motif-2$. Supplementary Fig.~\ref{fig:S10} and Fig.~\ref{fig:S11} describe the different regimes obtained by pacing one cell with $T = 300$~ms for $Motif-1$ and $Motif-3$. 

\begin{figure}[hbt!]
    \centering
    \includegraphics[width=0.75\textwidth]{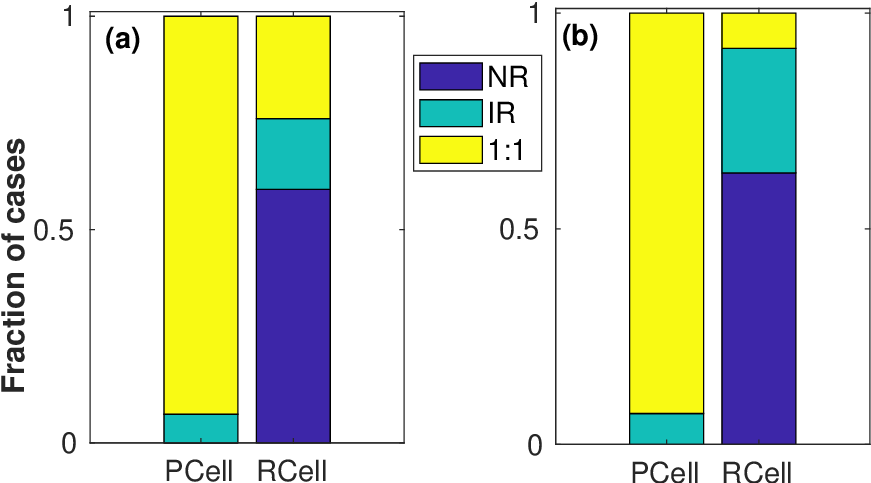}
    \caption{Effect of restitution. Fraction of occurrence of each dynamical regime, {\it viz.,} $NR$, $IR$ and $1:1$ summed over all pacing cyles and motifs at both {\it Pacing Cell} ($PCell$) and {\it Response Cell} ($RCell$) for {\it Shallow} (a) and {\it Steep} (b) parameter sets.} 
   \label{fig:EffectOfRestitution}
\end{figure}

\begin{figure}[hbt!]
    \centering
    \includegraphics[width=0.75\textwidth]{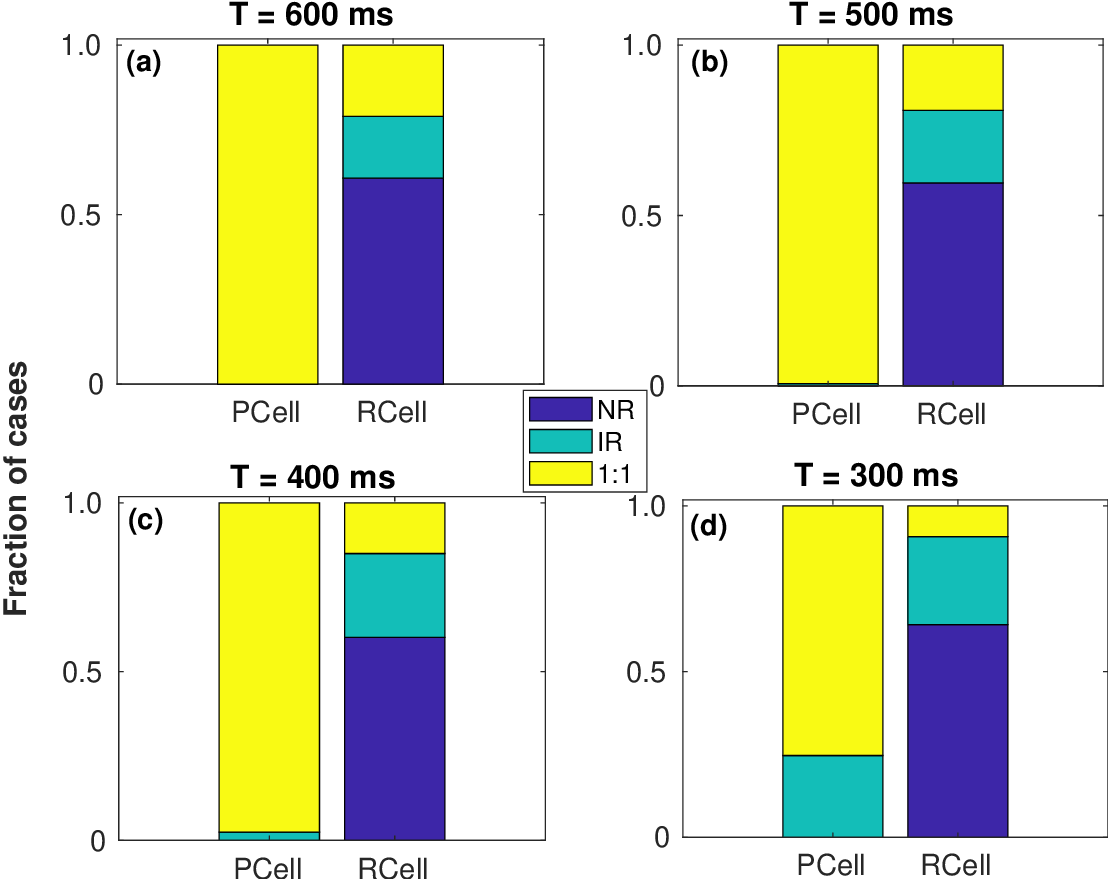}
    \caption{Effect of pacing period. Fraction of occurrence of each dynamical regime, {\it viz.,} $NR$, $IR$ and $1:1$ summed over all motifs and restitution types at both {\it Pacing Cell} ($PCell$)  and {\it Response Cell} ($RCell$) for pacing periods $T = 600$ ms, $T = 500$ ms, $T = 400$ ms and $T = 300$ ms.} 
   \label{fig:EffectOfPacingPeriod}
\end{figure}

\begin{figure}[hbt!]
    \centering
    \includegraphics[width=0.75\textwidth]{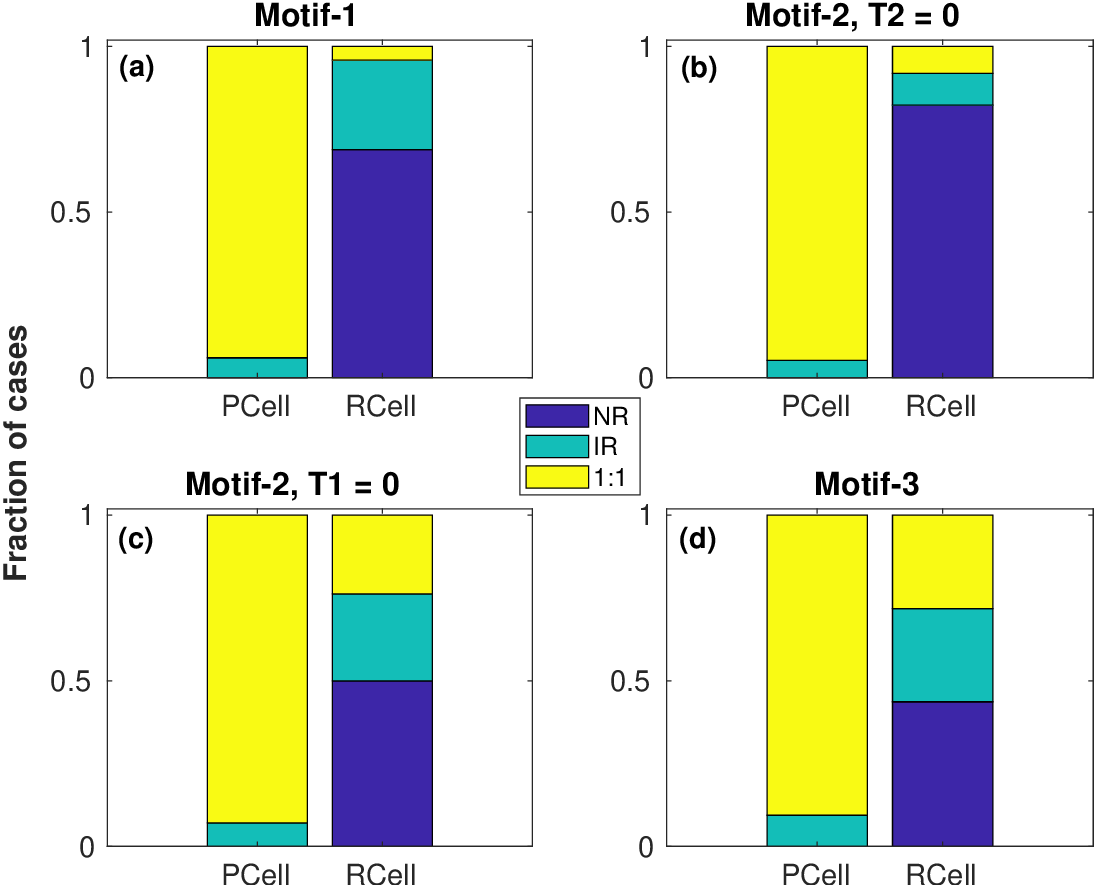}
    \caption{Effect of connection topology. Fraction of occurrence of each dynamical regime, {\it viz.,} $NR$, $IR$ and $1:1$ summed over all restitution types and pacing cycle lengths at both {\it Pacing Cell} ($PCell$)  and {\it Response Cell} ($RCell$) for $Motif-1$ (a), $Motif-2$ (b,c) and $Motif-3$ (d).}
    \label{fig:EffectOfMotif}
\end{figure}

The $2$\hyp{}parameter plots are useful in identifying the\\
different dynamical regimes and their relation to myocyte\hyp{}fibroblast coupling strength. However in order to determine the influence of each of the features ({\it viz.,} connection topology, myocyte parameters, pacing period) on the myocyte dynamics, we determined the fraction of instances for every regime, keeping one feature fixed at a time. In Fig.~\ref{fig:EffectOfRestitution} we plot the fraction of occurrence of each regime for {\it Steep} and {\it Shallow} parameters individually, summing across the other features for both {\it Pacing Cell} ($PCell$) and {\it Response Cell} ($RCell$). Similarly Fig.~\ref{fig:EffectOfPacingPeriod} and Fig.~\ref{fig:EffectOfMotif} describe the fraction of occurrence of the different regimes for the different pacing periods and motifs respectively. 

Based on Fig.~\ref{fig:EffectOfRestitution}, Fig.~\ref{fig:EffectOfPacingPeriod} and Fig.~\ref{fig:EffectOfMotif}, we can summarise that (i) Obtaining $1:1$ response was more difficult
with {\it Steep} parameters compared to {\it Shallow}. On the other hand {\it Steep} showed $IR$ for more parameters than {\it Shallow}. (ii) With increase in pacing period, there were fewer instances of $1:1$ and more cases of $IR$. For $T \leq 400$~ms the {\it Pacing Cell} can also elicit $IR$ instead of $1:1$ for some coupling strengths. (iii) $Motif-3$ with the smallest region of $NR$ and most instances of $1:1$ allowed for conduction over the largest range of parameters.
On the other hand $Motif-2$ ($T2 = 0$ in Fig.~\ref{fig:Motif}(b)) with the maximum instances of $NR$ %was the least efficient
had the smallest fraction of parameters that allowed any
conduction ($IR$ or $1:1$). $Motif-1$ had the least number of $1:1$ among all motifs but had a larger fraction of cases describing $IR$ than $Motif-2$. 

We also repeated our simulations with a more negative fibroblast resting membrane potential $V_{FR} = -49.0$~mV for $Motif-2$ and $Motif-3$. The results for a subset of these simulations are described in the Supplementary section.

\begin{figure}[!htb]
    \includegraphics[width=\textwidth]{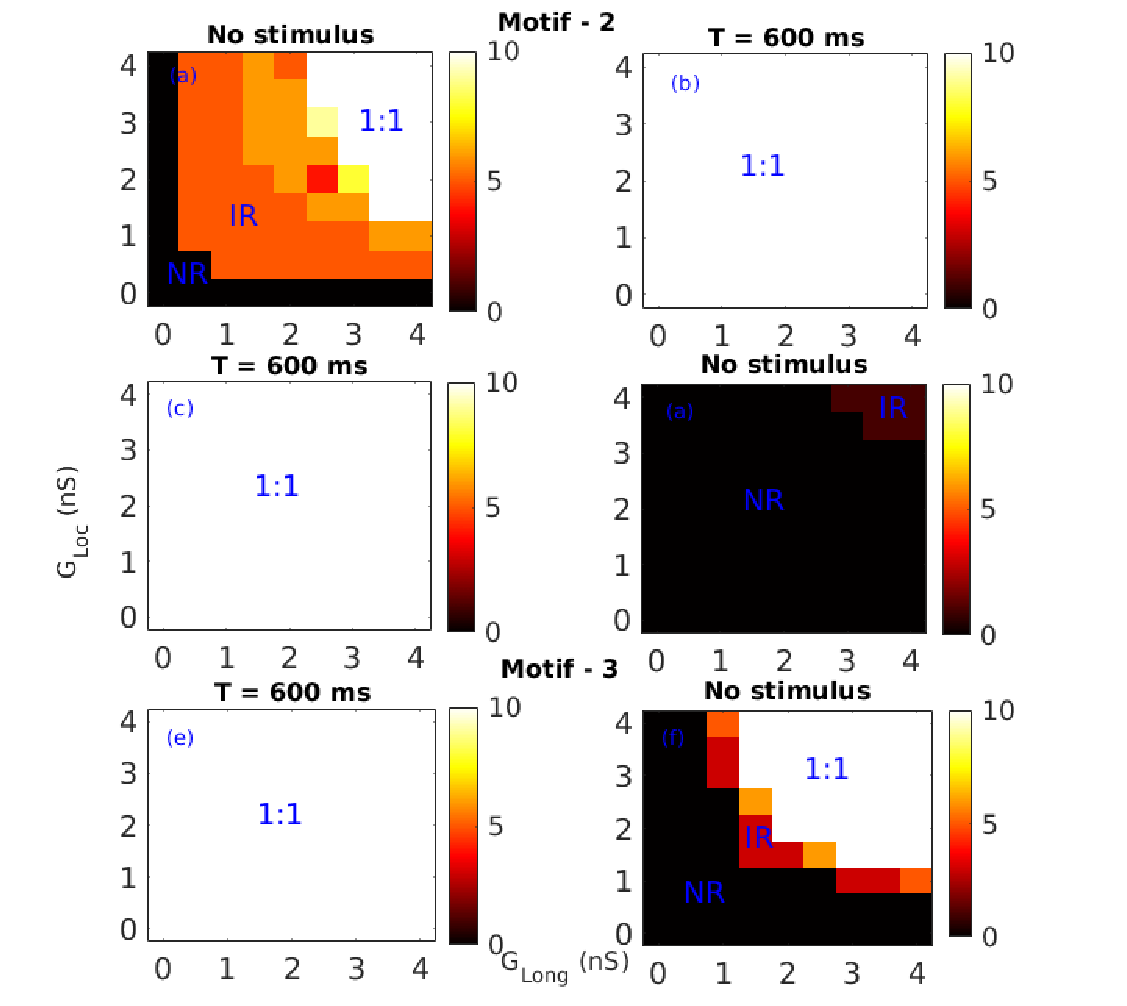}
    \caption{Identifying regimes for {\it Steep} parameters at $T = 600$ ms that initiate action potentials in non-stimulated cells for $V_{FR} = -49$ mV.}
    \label{fig:S13}
\end{figure}

\begin{figure}[!htb]
    \includegraphics[width=\textwidth]{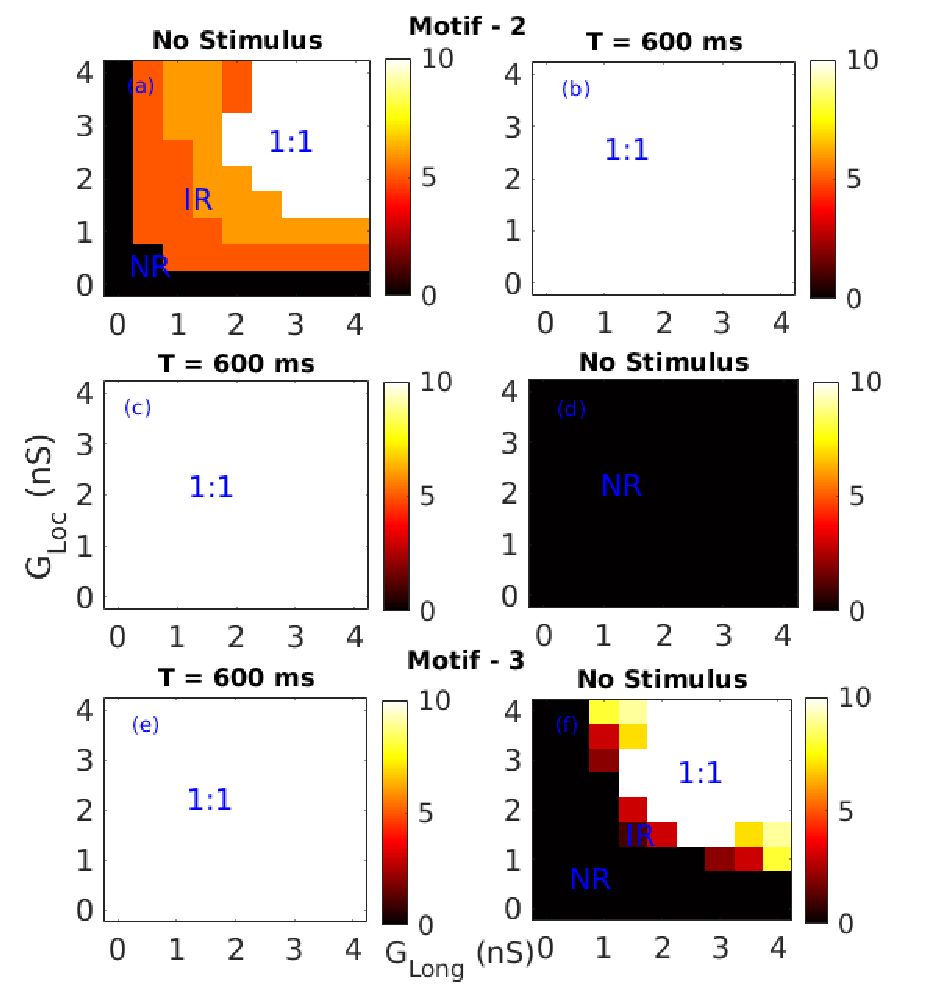}
    \caption{Identifying conductance parameters that initiate action potentials in non-stimulated myocyte with {\it Shallow} restitution for $V_{FR} = -49$ mV.}    
\label{fig:S15}
\end{figure}

A more negative fibroblast resting membrane potential resulted in a greater decrease in APD for both {\it Shallow} (Supplementary Fig.~\ref{fig:S14}) and {\it Steep} (Supplementary Fig.~\ref{fig:S12}) restitution parameters compared to a resting membrane potential of $-24.5$~mV. 
Supplementary figures Fig.~\ref{fig:S13} and Fig.~\ref{fig:S15} describe the $2$-parameter plots identifying the different dynamical regimes when only one myocyte was stimulated for {\it Steep} and {\it Shallow} regimes respectively. While the different dynamical regimes are qualitatively similar to those obtained in $V_{FR} = -24.5$mV, it can be observed that due to the more negative fibroblast resting potential a stronger coupling was required to elicit an action potential in the {\it Response Cell}. 

\section*{Discussion}
In this paper we have described a simple way to capture complex cellular dynamics that can occur due to the interaction between $2$ myocytes that are coupled only via fibroblasts. In the heart the absence of gap-junctional coupling between myocytes could be due to the presence of ablation lines, scars or fibrosis resulting in spatial and electrical separation of myocytes.
However under such conditions there is some evidence that conduction can still occur via fibroblasts across novel pathways coupling otherwise disconnected myocytes~\cite{Walker:2007,Kohl:2005,Rog:2016,Simon:2023}. As we have described in this paper, such connections can give rise to a range of dynamical behaviour including reduced APD, synchronization of repolarization in uncoupled myocytes, initiation of action potentials in resting cells and conduction delays. At the level of cardiac tissue such non-local coupling might possibly impact wave propagation and lead to reentry or conduction blocks.

We have considered $3$ different topological arrangements of $2$ myocytes coupled to either $1$ or $2$ fibroblast units and investigated their effect on myocyte dynamics.
We have chosen myocyte parameters corresponding to {\it Shallow} and {\it Steep} restitution slopes~\cite{tenTuss:2006} and for the fibroblast we have used the {\it MacCannell} ``active" fibroblast model~\cite{MacCannell:2007} with modifications made to obtain different resting membrane potentials~\cite{Jacquemet:2008}. Since the setup we have considered here is more likely in a heart tissue undergoing repair from either surgery, injury or disease we have modified the {\it MacCannell} model to simulate myofibroblasts by setting $C_f = 50$~pF. 

\begin{figure}[!htb]
    \includegraphics[width=\textwidth]{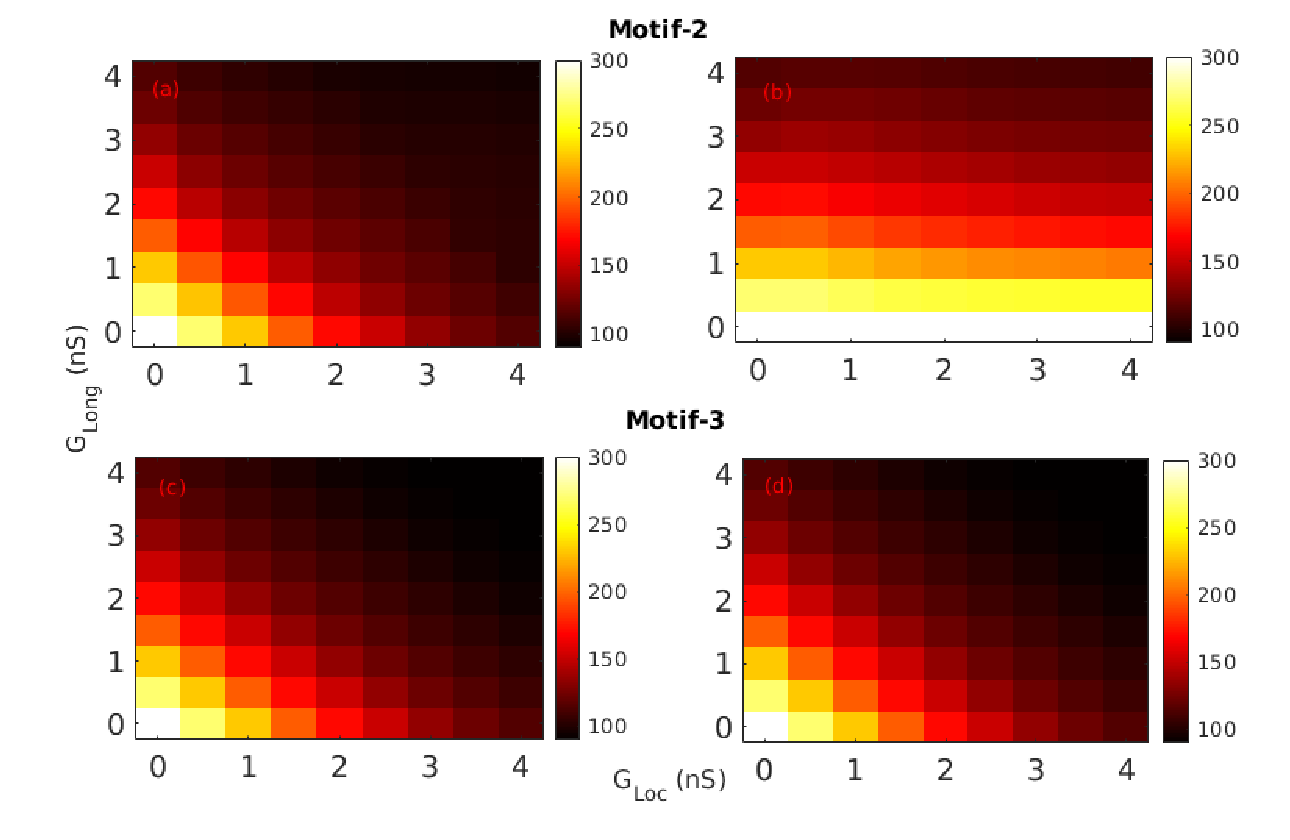}
    \caption{Effect of coupling motifs on APD for {\it Steep} parameters with $V_{FR} = -49$ mV.}
    \label{fig:S12}
\end{figure}

\begin{figure}[!htb]
    \includegraphics[width=\textwidth]{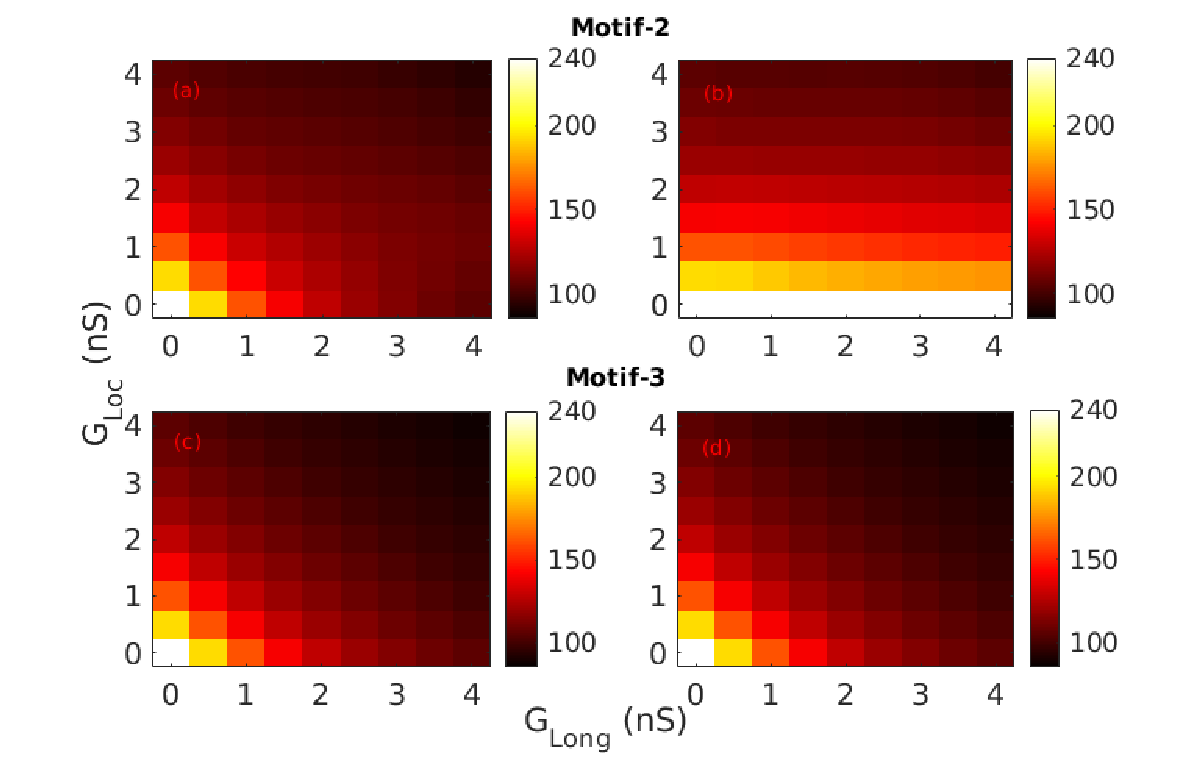}
    \caption{Effect of coupling motifs with {\it Shallow} cell parameters on APD for $V_{FR} = -49$ mV.}    
   \label{fig:S14}
\end{figure}

Irrespective of the connection topology, the primary effect of the coupling was a decrease in the APD of the myocyte with increase in coupling strength (Fig.~\ref{fig:APDMotif}) with the myofibroblast acting as a leaky capacitor~\cite{Nguyen:2012}. 
However the magnitude of the change in APD depended on the type of motifs and the myocyte parameters ({\it i.e.,} whether the restitution is {\it Steep} or {\it Shallow}). The decrease in APD is more pronounced for the case with {\it Steep} parameters (Fig.~\ref{fig:APDMotif}) compared to {\it Shallow} parameters (Supplementary Fig.~\ref{fig:S1}) across all the motifs.
Also the magnitude of the change in APD increased with the complexity of the connections with $Motif-3$ (the most complex and tightly coupled of the motifs considered) showing the greatest decrease in APD at maximum coupling.
Furthermore we observed that the decrease in APD also increases for a more negative resting membrane potential of the fibroblast (Supplementary figures Fig.~\ref{fig:S14} and Fig.~\ref{fig:S12}), an observation consistent with earlier studies~\cite{Nguyen:2012}. 	

\begin{figure}[!htb]
    \includegraphics[width=\textwidth]{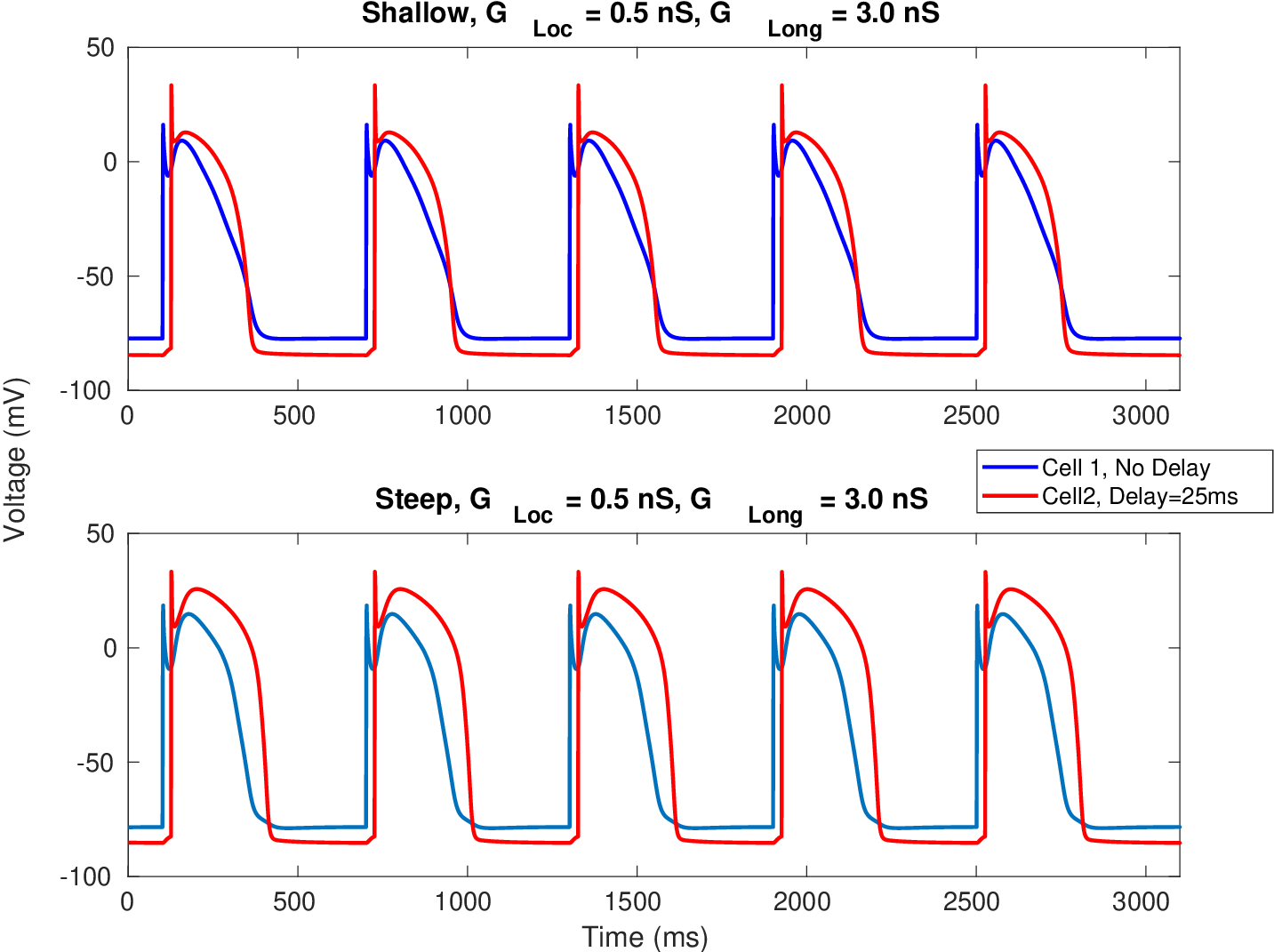}
    \caption{The time series of trans-membrane potential for both myocytes in $Motif-3$, with $myocyte-2$ stimulated $25$ ms after $myocyte-1$.}
    \label{fig:S5} 
\end{figure}

Since the two myocytes considered in the motifs are uncoupled and spatially separated, there is a time difference in the excitation of the two cells. This could be due to the difference in times that the cells are excited in the myocardium (i.e., the effect of conduction velocity across the heart tissue) and/or due to the effect of conduction delay in the propagation of current via fibroblasts. We first studied the case where the predominant delay is the time difference ($\tau_D$) between the stimulation of both myocytes (Fig.~\ref{fig:DelayMotif} and Supplementary Fig.~\ref{fig:S4}).
The delay in the stimulation of $myocyte-2$ breaks the symmetry of coupling and results in differential myocyte APD even for $Motif-1$ and $Motif-3$ (Supplementary Fig.~\ref{fig:S5}) for the action potential profiles for cells stimulated at different times). We observed that the sign and magnitude of the difference in APD of the two myocytes ($\Delta APD$) is sensitive to one or more of the features varied here {\it viz.,} connection topology, coupling strengths, delay ($\tau_D$) and the myocyte parameters ({\it Shallow} or {\it Steep}). For conditions of finite stimulus delay, $\tau_D$ acts as a proxy to the effect of conduction velocity across the myocardium.

Another type of delay that can occur in these systems is the time taken by the excitation to propagate from one myocyte to another purely via a fibroblast. In order to investigate the effect of this conduction we considered the case of $\tau_D \rightarrow \infty$. In practical terms this means that one of the cells in the motif is never stimulated externally for the duration of the simulation, while the other myocyte is stimulated periodically as before. Figure.~\ref{fig:APMotif3NoT2} (a-b) shows the $2$-parameter phase space characterizing the different dynamical regimes as a function of the gap-junctional coupling conductance.
Figure.~\ref{fig:APMotif3NoT2} (c-f) highlights the sensitive dependence of the myocyte dynamics on the coupling strength, with small changes in coupling parameters resulting in widely different myocyte responses. The idea of infinite delay in stimulation is especially useful to illustrate the dynamics of initiation of action potential in regions that are isolated from its neighbours except for conduction via fibroblasts. 

The intermittent sequence of action potentials observed in the $IR$ regime in the {\it Response Cell} (and in {\it Pacing Cell} at very rapid pacing) suggests a plausible dynamical mechanism that can result in conduction block and initiation of reentry in tissue. Fig.~\ref{fig:EffectOfRestitution}, Fig.~\ref{fig:EffectOfPacingPeriod} and Fig.~\ref{fig:EffectOfMotif} summarize the role of the individual features in determining the dynamical regimes. While steep restitution and rapid pacing favour complex or irregular dynamics in both the {\it Response} and {\it Pacing Cells}, the motif structure and the location of the stimulated myocyte (with respect to fibroblasts connected to it) also play a critical role in determining the dynamical regime. In particular $Motif-1$ (where one fibroblast couples to two uncoupled myocytes) with its very low number of $1:1$ response is a very plausible connection topology in tissue that can give rise to conduction blocks and reentry.

While heterocellular coupling between myofibroblasts and myocytes have been reported to initiate ectopic activity {\it in vitro}~\cite{Miragoli:2007}, to the best of our knowledge this is the first {\it in silico} study to systematically investigate the different features that can potentially influence the myocyte dynamics when coupled purely via fibroblasts. We hypothesised that non-local coupling could potentially result in the initiation of action potentials in the quiescent myocyte and identified the parameters that described the various dynamical regimes possible for different connection topology.

Motifs are a simple prototype to investigate the effect of long-distance connections between uncoupled myocytes connected only via fibroblasts.
Many studies have looked at the effect of fibroblast distributions in simulated tissue on wave dynamics ~\cite{Kazbanov:2016,Sridhar:2017,tenTuss:2007}.
The idea of fibrotic and functional clusters~\cite{Alonso:2013,Panfilov:2021} have been developed based on percolation theory to investigate the interaction of wave-propagation with fibrosis. Interstitial fibrosis is associated with non-ischemic cardiomyopathy and has been modelled as infinitesimal splits in a finite element mesh ~\cite{Balaban:2020}.
Machine learning algorithms have been employed to understand the effect of local fibrosis patterns especially at border zones~\cite{Zahid:2016}. More recently homogenisation techniques have been applied to model fibrosis as spatially repeating structures~\cite{Gokhale:2018}, or by using graph theoretical~\cite{Farquhar:2022} and volume averaging approaches~\cite{Lawson:2023} to incorporate microscopic structures into a macro-scale problem. 
 
Our methodology in this study differs significantly from the methods used in the above papers. We have adopted a bottom-up approach where we have developed simple structural motifs that can in principle be scaled to build scar boundary zones. While motifs have been used extensively in other areas~\cite{Milo:2002,Song:2005,Alon:2007}, to the best of our knowledge this is the first paper that uses the idea of motifs to describe the various dynamics arising from myocyte-fibroblast interaction.
Our approach is especially suitable to study the effect of fibroblast connections that form across ablation lines. Fibroblasts that are involved in the repair of the surgically ablated zones could enable conduction across the separated regions~\cite{Rog:2016}. The motifs described here provide a possible structural mechanism to couple disconnected regions in tissue.  

We conclude by stating the limitations of our study and the scope of future work. While the motifs developed here are a useful prototype to simulate non-local coupling in cells across ablation line or fibrotic regions, they do not account for the effect of electrotonic diffusion, which acts to smooth wavefronts and reduce the effect of local variations in topology. So in as much as motifs can be a useful tool to quickly explore dynamics while varying several factors their maximal utility would be realised when they are used to build scar tissue boundaries and simulate conduction across non-conducting ablation lines in heart tissue. While we have considered homogeneous myocytes in our motifs, myocyte properties are heterogeneous in diseased hearts. Expanding the motif prototype to $2D$ tissue and incorporating heterogeneous myocytes in the motifs are two areas of future work. In our simulations we fixed the number of fibroblasts in a unit ($N_f$) connected to a myocyte to be $4$. But the number of fibroblasts coupled to a myocyte is an important factor that affects the action potential and can be systematically varied to investigate its effect on wave propagation~\cite{Mortensen:2021}. 
Lastly an important limitation of this paper is that we have only simulated the electrophysiology of myocytes and not the mechanics of their contraction and relaxation. Mechanical contractions and the resultant change in tissue geometry have significant effect on wave-propagation. Another factor that influences wave\hyp{}propagation is the mechano\hyp{}electric feedback. These are aspects that will be incorporated in future studies.
\bibliographystyle{unsrt}
\bibliography{References}

\section*{Acknowledgements}
SS and RHC would like to acknowledge EPSRC EP/T017899/1 The SofTMech Statistical Emulation and
Translation Hub for funding SS. SS and RHC would like to thank Prof Godfrey Smith and Prof Radostin Simitev for useful suggestions and comments. SS would also like to thank Prof Sitabhra Sinha for useful discussions.

\section*{Supplementary Information}
\begin{figure}[!htb]
    \includegraphics[width=\textwidth]{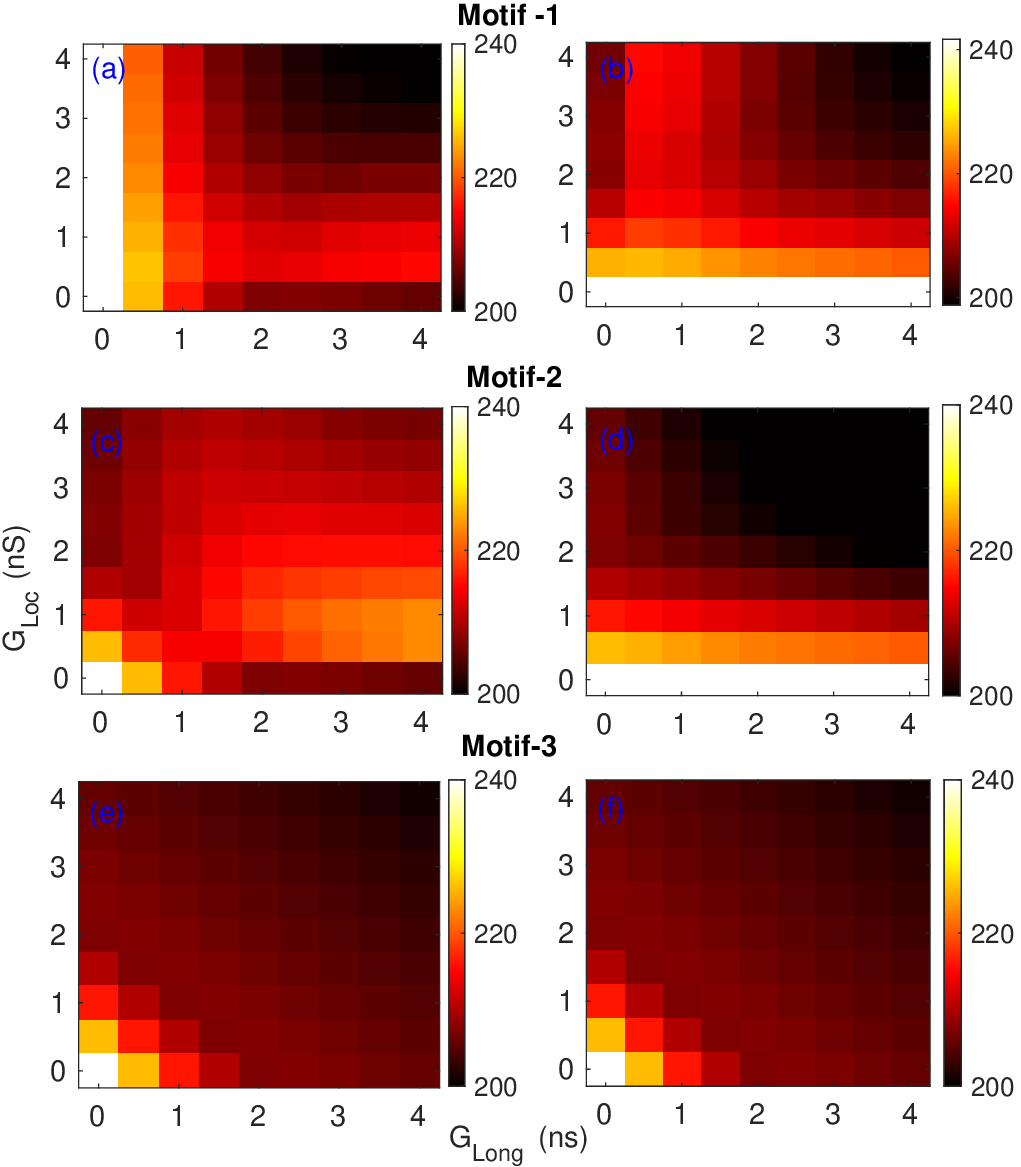}
    \caption{$2$-parameter plots for all the $3$ motifs describing the effect of the coupling strengths $G_{Loc}, G_{Long}$ on the APD for both the myocytes with Shallow parameters.}
    \label{fig:S1}
\end{figure}

\begin{figure}[!htb]
    \includegraphics[width=\textwidth]{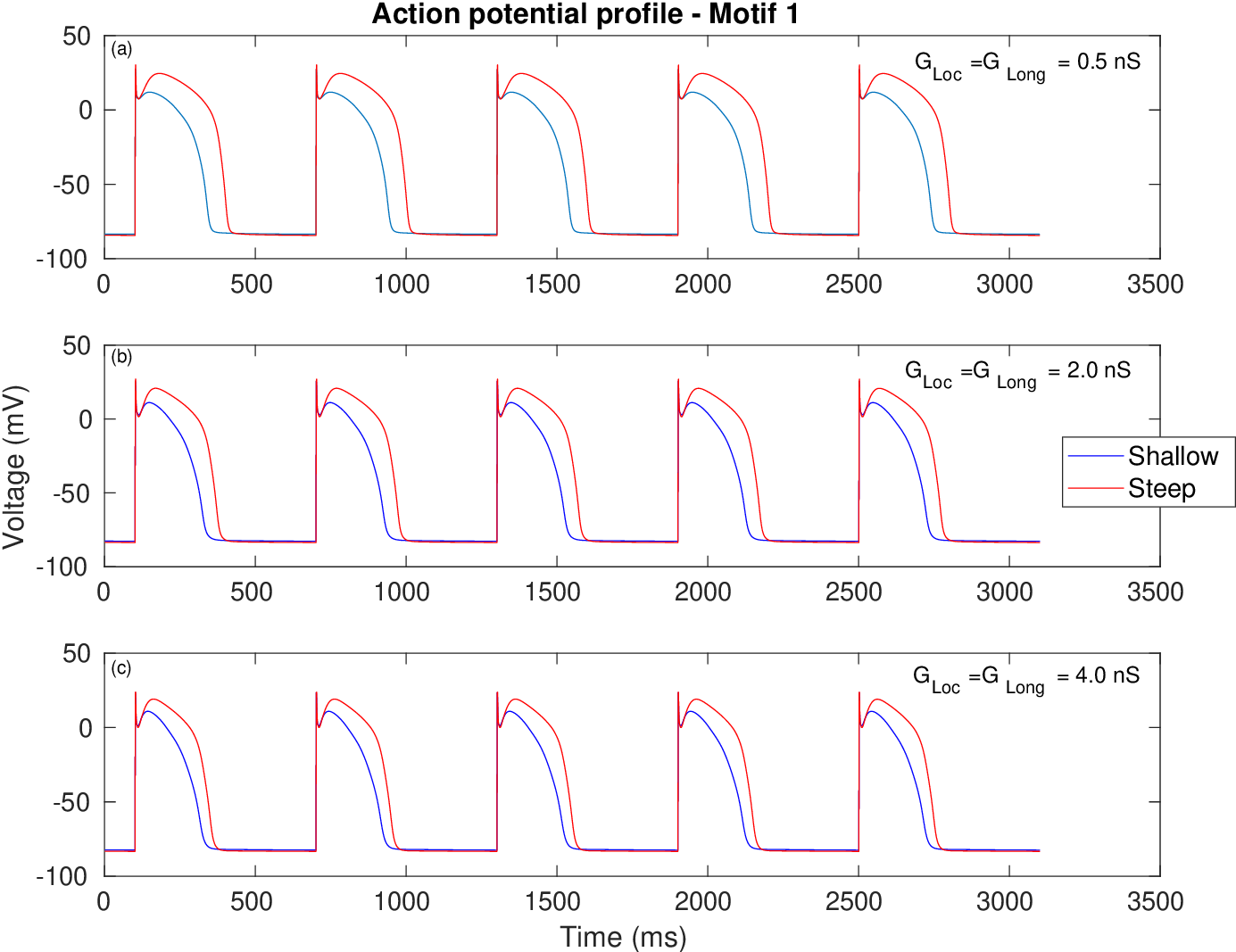}
    \caption{Action potential profiles for both myocytes in $Motif-1$ while pacing the cells at $T=600$ ms at $G_{Loc} = G_{Long} = 0.5$ nS (a), $2.0$ nS (b) and $4.0$ nS (c) respectively. The blue and red traces correspond to {\it Shallow} and {\it Steep} parameter sets.}
    \label{fig:S2}
\end{figure}

\begin{figure}[!htb]
    \includegraphics[width=\textwidth]{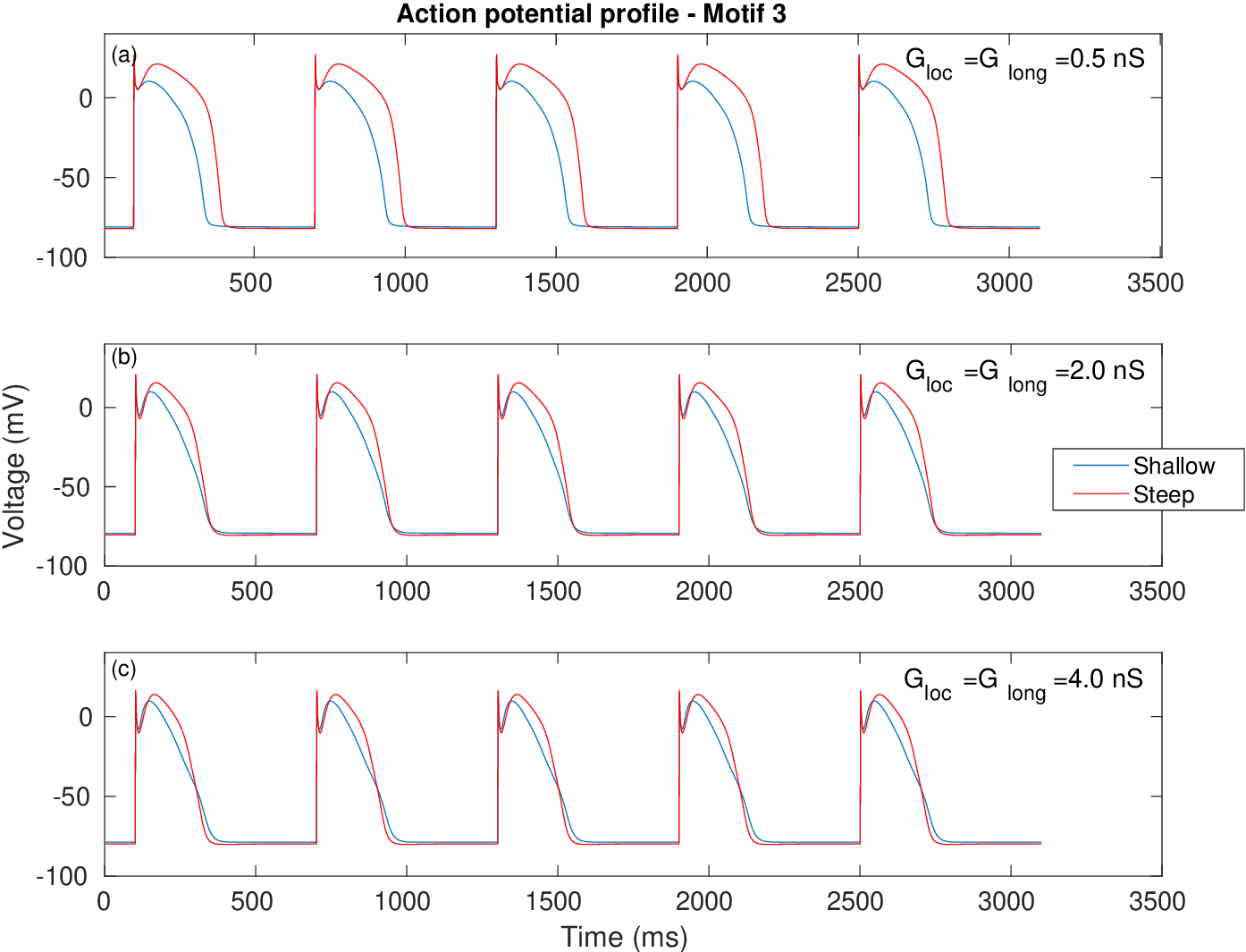}
    \caption{Action potential profiles for both myocytes in $Motif-3$ while pacing the cells at $T=600$ ms at $G_{Loc} = G_{Long} = 0.5$ nS (a), $2.0$ nS (b) and $4.0$ nS (c) respectively. The blue and red traces correpsond to {\it Shallow} and {\it Steep} parameter sets.}
    \label{fig:S3} 
\end{figure}


\end{document}